\documentstyle[12pt]{article}
\catcode`\@=11
\def\newsymbol#1#2#3#4#5{\let\next@\relax%
 \ifnum#2=\@ne\else%
 \ifnum#2=\tw@\let\next@\msyfam@\fi\fi%
 \mathchardef#1="#3\next@#4#5}
\def\mathhexbox@#1#2#3{\relax%
 \ifmmode\mathpalette{}{\m@th\mnnathchar"#1#2#3}
 \else\leavevmode\hbox{$\m@th\mathchar"#1#2#3$}\fi}
\font\tenmsy=msbm10
\font\sevenmsy=msbm7
\font\fivemsy=msbm5
\newfam\msyfam
\textfont\msyfam=\tenmsy
\scriptfont\msyfam=\sevenmsy
\scriptscriptfont\msyfam=\fivemsy
\edef\msyfam@{\hexnumber@\msyfam}
\def\Bbb#1{\fam\msyfam\relax#1}
\catcode`\@=\active

\newtheorem{theorem}{Theorem}[section]
\newtheorem{proposition}[theorem]{Proposition}
\newtheorem{lemma}[theorem]{Lemma}
\newtheorem{corollary}[theorem]{Corollary}

\newtheorem{example}[theorem]{Example}
\newtheorem{remark}[theorem]{Remark}

\newcommand{\eq}[1]{\begin{equation}\label{#1}}
\newcommand{\en}{\end{equation}}
\newcommand{\eqn}{\begin{eqnarray*}}
\newcommand{\enn}{\end{eqnarray*}}
\newcommand{\eqnn}{\begin{eqnarray}}
\newcommand{\ennn}{\end{eqnarray}}
\newcommand{\proof}{{\noindent \it Proof:\ }}
\newcommand{\qed}{\hfill $\Box$\par\medskip}
\newcommand{\BR}{{{\Bbb R}^{\nu}}}

\newcommand{\bi}{\begin{description}}
\newcommand{\ei}{\end{description} }
\newcommand{\CC}{{{\Bbb C}}}

\newcommand{\RR}{{\Bbb R}}
\newcommand{\ez}{\epsilon_0} 
\newcommand{\fox}{f_{\omega \lambda}^x} 
\newcommand{\foo}{f_{\omega \lambda^0}} 

\newcommand{\limm}{\lim_{m\rightarrow\infty}}
\newcommand{\limt}{\lim_{t\rightarrow\infty}}

\newcommand{\kak}[1]{(\ref{#1})}
\newcommand{\LR}{{L^2(\BR)}}
\newcommand{\LRR}{{L^2(\RR^3)}}

\newcommand{\LM}{{L^2(M)}}

\newcommand{\fff}{{{\cal F}_{\rm b}}}
\newcommand{\FFF}{{{\cal F}}}
\newcommand{\FFFg}{{{\cal F}_{\rm SB}}}
\newcommand{\FFFp}{{{\cal F}_{\rm PF}}}
\newcommand{\FFFn}{{{\cal F}_{\rm N}}}
\newcommand{\gsb}{H_{\rm SB}}
\newcommand{\gsbz}{H_{\rm SB,0}}
\newcommand{\gsbi}{H_{\rm SB,I}}
\newcommand{\pfi}{H_{\rm PF,I}}
\newcommand{\pfz}{H_{\rm PF,0}}
\newcommand{\n}{H_{\rm N}}
\newcommand{\nr}{H_{\rm N}^{\rm reg}}
\newcommand{\nri}{H_{\rm N,I}^{\rm reg}}
\newcommand{\nrii}{\tilde{H}_{\rm N,I}^{\rm reg}}
\newcommand{\nz}{H_{\rm N, 0}}
\newcommand{\Ni}{H_{\rm N, I}}
\newcommand{\tgsb}{T_{\rm SB}}
\newcommand{\tpf}{{T_{\rm PF}}}
\newcommand{\tn}{T_{\rm N}}
\newcommand{\tnr}{T_{\rm N}^{\rm reg}}
\newcommand{\jjj}{\sum_{j=1}^J}

\newcommand{\hFFF}{{\cal F}}
\newcommand{\ffff}{{\cal F}_{\rm b,0}}
\newcommand{\fffi}{{\cal F}_\infty} 

\newcommand{\f}{^{-1}}
\newcommand{\PF}{H_{\rm PF}}
\newcommand{\lk}{\left(}
\newcommand{\rk}{\right)}
\newcommand{\lkk}{\left\{}
\newcommand{\rkk}{\right\}}
\newcommand{\lan}{\langle}
\newcommand{\ran}{\rangle}

\newcommand{\bl}[1]{\begin{lemma}\label{#1}}
\newcommand{\el}{\end{lemma}}
\newcommand{\bc}[1]{\begin{corollary}\label{#1}}
\newcommand{\ec}{\end{corollary}}
\newcommand{\bt}[1]{\begin{theorem}\label{#1}}
\newcommand{\et}{\end{theorem}}
\newcommand{\bp}[1]{\begin{proposition}\label{#1}}
\newcommand{\ep}{\end{proposition}}
\newcommand{\br}[1]{\begin{remark}\label{#1}}
\newcommand{\er}{\end{remark}}
\newcommand{\ha}{a} 
\newcommand{\add}{a^{\ast}}

\newcommand{\www}{{\cal K}}
\newcommand{\kkk}{{\cal K}}
\newcommand{\k}[1]{{\otimes_s^{#1}\kkk}}
\newcommand{\ass}{a^\sharp}

\newcommand{\ov}[1]{\overline{#1}}

\newcommand{\gr}{\varphi_{\rm g}}

\newcommand{\pg}{P_{\rm g}}
\newcommand{\half}{\frac{1}{2}}
\newcommand{\han}{{1/2}}
\newcommand{\hhh}{{\cal H}}

\newcommand{\hi}{H_{\rm I}}
\newcommand{\ihh}{K_{\rm I}}
\newcommand{\wh}{H} 
\newcommand{\dg}{d\Gamma}
 
\renewcommand{\i}{1}
\newcommand{\ii}{\i\otimes }
\newcommand{\la}{\lambda}
\newcommand{\fs}{S} 
\newcommand{\s}{\sigma}
\newcommand{\ks}{k_{\sigma(j)}}
\newcommand{\ff}[1]{f_{\s(#1)}}
\newcommand{\nn}{^{(n)}}

\newcommand{\m}[1]{{\cal M}_{#1}}
\newcommand{\mm}{\m{-\han}\cap\m{[n/2]}}
\newcommand{\mmm}{\m{-\han}\cap\m 0}

\newcommand{\ad}[3]{{\rm ad}_{#1}^{#2}(#3)}
\newcommand{\ada}{{\rm ad}_A} 
\newcommand{\adxy}{{\rm ad}_{\D+\beta V}} 
\newcommand{\adx}{{\rm ad}_{\D}} 
\newcommand{\ady}{{\rm ad}_V} 
\newcommand{\J}{\sum_{j_1,..,j_\ell=1}^\nu} 

\newcommand{\T}{K}
\newcommand{\cc}[2]{\lk\!\!\!\begin{array}{c}#1\\ #2 \end{array}\!\!\!\rk}
\renewcommand{\sl}{\backslash\!\!\!\!}

\newcommand{\oh}{\widehat{\wh}}
\newcommand{\cd}{{\cal C}_0}
\newcommand{\ccd}{{\cal C}_n}
\newcommand{\di}{{\cal D}_\infty}
\renewcommand{\aa}[3]{{\rm ad}_{#1}^{#2}(#3)}
\newcommand{\hz}{K_0} 
\newcommand{\avv}{\hat A} 
\newcommand{\p}[1]{p_{j_{#1}}}
\newcommand{\pp}[1]{p_{i_{#1}}}
\newcommand{\D}{(-\Delta)} 
\renewcommand{\S}{{\cal S}(\BR)} 
\newcommand{\gk}{g^{(k)}_{j_1\cdots j_\ell}}
\newcommand{\g}[1]{g^{(#1)}}
\newcommand{\q}[1]{T(k_{#1})^\ast e^{it_{#1}\oh} \cdots T(k_1)^\ast e^{it_1\oh}}
\newcommand{\qgsb}[1]{\tgsb(k_{#1})^\ast e^{it_{#1}\wgs} \cdots \tgsb(k_1)^\ast e^{it_1\wgs}}

\newcommand{\qm}[1]{T(k_{\s({#1})})^\ast e^{it_{#1}(\oh+\sum_{j={#1}}^n\fs(\ks))} 
\cdots T(k_{\s(1)})^\ast e^{it_1(\oh+\sum_{j=1}^n\fs(\ks))}}
\newcommand{\qgsbm}[1]
{\tgsb(k_{\s({#1})})^\ast e^{it_{#1}(\wgs+\sum_{j={#1}}^n\omega(\ks))} 
\cdots \tgsb(k_{\s(1)})^\ast e^{it_1(\wgs+\sum_{j=1}^n\omega(\ks))}}
\newcommand{\wgs}{\widehat{H}_{\rm SB}}
\newcommand{\nnn}{\nonumber}

\makeatletter
\@addtoreset{equation}{section}
\makeatother

\title
{Regularities of ground states of quantum field models}

\author{
Asao Arai\thanks{Department of Mathematics,  Hokkaido
University, Sapporo 060-0810, Japan.},
Masao Hirokawa\thanks {Department of Mathematics, Faculty of Science,
Okayama University, Okayama 700-8530, Japan.}
and  Fumio Hiroshima\thanks{
Department of Mathematics and Physics,
Setsunan University,   Osaka 572-8508, Japan.}
}
\date{\today}
\begin{document}
\pagestyle{myheadings}
\markboth{Regularities of ground states}{Regularities of ground states}

\setlength{\baselineskip}{16pt}
\maketitle

\begin{abstract}
Regularities  and higher order regularities  of ground states of 
quantum field models are investigated through the fact that asymptotic annihilation operators 
vanish ground states. 
Moreover a sufficient condition for the absence of a ground state is given.
\end{abstract}


\section{Introduction}
A basic object in a quantum field model is a ground state
of it which is defined to be an eigenvector of the Hamiltonian
$H$ (a self-adjoint operator on a Hilbert space) 
 of the model with eigenvalue equal to
the infimum $E(H)$ of the spectrum of $H$, provided that
$E(H)$ is an eigenvalue of $H$. 
In this paper
we investigate regularities of ground states of 
quantum field models. Here  we mean by regularities
properties of  what class of subspaces 
ground states belong to, including absence of ground states
in certain subspaces.

 As is well known, 
existence of a ground state of a quantum field model 
may depend on
properties of objects contained in its Hamiltonian
such as a one-particle energy function and cutoff functions.
In particular, 
it is subtle in the case where the quantum field is  massless 
 (\cite{ahh,a2,h00,h0,h1,h2,lomisp} 
and references therein), being related to the so-called
infrared divergence \cite{pf}.
From this point of view, it is important
to characterize regularities of ground states, in particular,
absence of them,
in terms of objects contained in the Hamiltonian
of the model under consideration, as generally as possible.
This is the main motivation of this work.
Preliminary work concerning this theme
was done in a previous paper \cite{ahh}, where the absence of
ground states of an abstract and general model, called the
{\it generalized spin boson} (GSB) model, was discussed.
In the present work we extend  results
obtained in \cite{ahh} to a more general class of quantum field
models, establishing new criteria for  regularities of
ground states.

This  paper is organized as follows. In Section 2 we define
the quantum field model to be considered. We prove
general results on regularities of ground states of the model.
Section 3 is concerned with absence of ground states of the model.
In Section 4 we consider 
higher order regularities
of ground states, where 
higher order regularities means properties
that ground states belong to smaller subspaces indexed
by powers of a nonnegative self-adjoint operator.
In Section 5 we apply the general results established in the
previous sections to the GSB model
and obtain results which extend those in \cite{ahh}.
In the last section we give some remarks on other
quantum field models in view of the present work.

\section{Regularities  of ground states: general aspects}
\subsection{Fock spaces and second quantizations}
Let $\www$ be a separable Hilbert space over complex field $\CC$, 
and 
$\otimes_s^n\www$ denote the $n$-fold symmetric tensor product of
$\www$ with $\otimes_s^0\www:=\CC$.
The norm and the scalar product on
Hilbert space ${\cal X}$ are denoted by $
\|f\|_{\cal X}$ and $(f,g)_{\cal X}$, $f,g\in{\cal X}$, 
respectively, 
where $(f,g)_{\cal X}$ is anti-linear in $f$ and linear in $g$.
The norm of bounded operator from ${\cal X}$ to a Hilbert space 
${\cal Y}$  is denoted by 
$\|X\|_{{\cal X}\rightarrow{\cal Y}}$ and the domain of unbounded operator $Y$ is by $D(Y)$. 
The Boson Fock space over $\www$ is defined by
$$\fff(\www):=
\bigoplus_{n=0}^\infty [\otimes_s^n \www]=
\{ \Psi=\{\Psi^{(n)}\}_{n=0}^\infty | 
\|\Psi\|_{\fff(\kkk)}^2:
=\sum_{n=0}^\infty\|\Psi^{(n)}\|_\k{n}^2<\infty\}.$$
The Fock vacuum is defined by
$$\Omega:=\{1,0,0,...\}\in\fff(\kkk)$$
and the finite particle subspace of $\fff(\kkk)$ by
$$\ffff(\www):=\lkk \{ \Psi^{(n)} \}_{n=0}^\infty\in\fff(\www) |
\Psi^{(n)}=0
\mbox{ for all } n\geq n_0 \mbox{ with some}\   n_0\rkk.$$
It is known that $\ffff(\www)$ is dense in $\fff(\www)$.
The annihilation operator $a(f)$ with $f\in\www$ is defined to 
be a densely defined closed operator on $\fff(\www)$ whose 
adjoint is given by
$$(a(f)^\ast\Psi)^{(n)}:=
\sqrt n S_n(f\otimes\Psi^{(n-1)}),\ \ \ \Psi\in  D(a(f)^\ast),$$
 where $S_n$ denotes the symmetrization operator on $\otimes^n\www$, i.e.,
$S_n(\otimes^n\www)=\otimes_s^n\www$.
We note that $a(f)$ is anti-linear in $f$ and $\add(g)$ linear in $g$.
The opertors $a(f)$ and $\add(f)$ leave $\ffff(\kkk)$ invariant and 
satisfy canonical commutation relations on
$\ffff(\kkk)$:
\eqnn
\label{c1}
&& [a(f),\add(g)]=(f,g)_\kkk,\\
\label{c2}
&&[a(f), a(g)]=0,\\
\label{c3}
&&[\add (f), \add (g)]=0.
\ennn
Since $a(f)$ and $\add(f)$ are closable  operators,
their closed extensions are denoted 
by the same symbols, respectively.
Define $\ffff^{\cal D}(\www)$ with subspace ${\cal D}\subset \kkk$ by the finite linear hull of
$$\lkk \add(f_1)\cdots \add(f_n)\Omega, \Omega| f_j\in D,j=1,...,n, n\geq
1\rkk.$$
It is known that $\ffff^D(\www)$ is dense in $\fff(\www)$ if $D$ is dense in ${\cal K}$. 
Let $C$
be a closed  operator  on $\www$.
Define $\dg_n(C):\otimes_s^n\kkk\rightarrow \otimes_s^n\kkk$ by
$$\dg_n(C):=\sum_{j=1}^n\underbrace{
\i\otimes\i \cdots \i\otimes \stackrel{j}{\breve C}\otimes\i \cdots\i
\otimes\i}_n,\ \ \ n\geq 1,
$$
and $\dg_0(C):\CC\rightarrow \CC$ by
$$\dg_0(C) z=0,\ \ \ z\in\CC.$$
The second quantization  of $C$ is
the operator
defined by
$$\dg (C):=
\bigoplus_{n=0}^\infty \dg_n(C)$$
with $D(\dg (C)):=\ffff^{D(C)}(\kkk)$.
Note that
\eqn
&& \dg (C)\Omega=0\\
&&
\dg (C)\add(f_1)\cdots \add(f_n)\Omega=
\sum_{j=1}^n \add(f_1)\cdots \add(C f_j)\cdots \add(f_n)\Omega, \\
&&\hspace{7cm} \ \  \ f_j\in  D(C),\ \ \ j=1,...,n.
\enn
For a nonnegative self-adjoint operator $K$,
$\dg (K)$ is a nonnegative essentially self-adjoint operator.
We denote its self-adjoint extension
by the same symbol, $\dg (K)$. 
$\dg(\i)$ is refered to as the number operator, which is denoted by 
$$ N:=\dg(\i).$$
It is known that, for all $\Psi\in D(\dg (K)^\han)$ and $f\in D(K^{-\han})$,
\eqnn
\label{2}
\|a(f)\Psi\|_{\fff(\www)}^2  &\leq & \| K^{-\han}f\|_{\cal K}^2 \|\dg (K)^\han\Psi\|_{\fff(\www)}^2,\\
\label{3}
\|\add (f)\Psi\|_{\fff(\www)}^2 &\leq & \| K^{-\han}f\|_{\cal K}^2 \|\dg
(K)^\han\Psi\|_{\fff(\www)}^2+\|f\|_{\cal K}^2\|\Psi\|_{\fff(\www)}^2.
\ennn
\bl{arai}
Let $0\leq \epsilon<1$ and $n$ be a non-negative integer.
Let $f\in D(K^{-1/2})\cap D(K^{n+1})$. Then
$\ass(f)$ maps $D(\dg(K)^{n+\epsilon+1/2})$ into $D(\dg(K)^{n+\epsilon})$.
In particular, it follows that
for $\Psi\in D(\dg(K)^{n/2})$ and $f_j\in D(K^{-\han})\cap D(K^{[n/2]})$,
$j=1,...,n$,
$$\Psi\in D(\ass(f_1)\cdots \ass(f_n)),$$
where
$\ass$ denotes $a$ or $\add$, and $[n/2]$ the integer part of $n/2$.
Moreove it follows that 
\eqn 
&& [\dg(K), a(f)]\Psi=-a(Kf),\ \ \ \Psi\in D(\dg(K)^{3/2}),\ \ \ f\in D(K),\\
&& [\dg(K), \add (f)]\Psi=\add (Kf),
\ \ \ \Psi\in D(\dg(K)^{3/2}),\ \ \ f\in D(K).
\enn \el
\proof
See \cite[Lemmas 2.3 and 2.5]{a1}. \qed

\subsection{Total Hamiltonians}
Let $\hhh$ be a Hilbert space over $\CC$. 
Then one can make the Hilbert space
$$\FFF:=\hhh\otimes\fff(\www).$$
Let $A$ be a self-adjoint operator  bounded from below on $\hhh$ and $S$ a
nonnegative self-adjoint operator on $\hhh$.
The decoupled Hamiltonian 
$$H_0:=A\otimes \i +\i \otimes \dg(S)$$
is self-adjoint on
$$D(H_0):=D(A\otimes \i )\cap D(\i \otimes \dg (S))$$
and bounded from below.
Let $\hi$ be a symmetric operator on $\FFF$ such that
$D(H_0)\subset D(\hi)$.
The total Hamiltonian under consideration is
the symmetric operator
$$H:=H_0+g \hi$$
on $\FFF$, where $g\in\RR$ is a coupling constant.
Assumption (A.1) is as follows. 
\bi
\item[(A.1)] There exist constants $a\geq 0$ and $b\geq 0$ such that
$$\|\hi \Psi\|\leq a\| H_0\Psi\|+b\|\Psi\|,\ \ \ \Psi\in D(H_0).$$
\ei
\bp{ko1}
Suppose (A.1). Then for $g$ with $|g|<1/a$,
$H$ is self-adjoint on $D(H_0)$ and bounded from below. 
Moreover it is  essentially self-adjoint on any core of
$H_0$.
\ep
\proof
It follows from the Kato-Rellich theorem \cite{rs2}.
\qed

It is known that there exist
a finite measure space $\lan M, \mu\ran$,
a nonnegative measurable function $\fs$ on $\lan M, \mu\ran$ and
a unitary operator
$U:\www\rightarrow \LM:=L^2(M,d\mu)$
such that
\bi
\item[(1)] $\Psi\in D(S)$ if and only if $\fs(\cdot)(U\Psi)(\cdot)\in \LM$,
\item[(2)] $(USU\f\Psi)(k)=\fs(k)\Psi(k)$ for $\Psi\in UD(S)$ for almost every  
$k\in M$.
\ei
Hence, without loss of generality, we can take $\www$ to be an 
$L^2$-space. Thus, in what follows, we set 
$$\www=L^2(M,d\mu),$$ 
where we do not assume that 
$\mu$ is a finite measure, but $\sigma$-finite measure, 
and take $S$ to be a multiplication operator on $\www$ by a non-negative function $S(k)$ such that 
$$0<S(k)<\infty, \ \ \  \mu-a.e. k.$$ 
We set
$$\m{m}:=D(\fs^m).$$

\subsection{Regularities  of ground states}
We denote $\inf \sigma(K)$ by $E(K)$ for a self-adjoint operator $K$. 
Let us assume that a ground state $\gr$ of $\wh $ exists.
Then
$$\wh  \gr=E(\wh )\gr.$$
We fix a normalized ground state $\gr$ of $\wh $, i.e., $\|\gr\|_\hFFF=1$.
Let $B$ and  $C$ be operators on a Hilbert space ${\cal W}$.
We define a quadratic form $[B, C]_W^{\cal D}$ with form domain 
${\cal D}\times{\cal D}$
such that 
$${\cal D}\subset D(B^\ast)\cap D(B)\cap D(C^\ast)\cap D(C),$$
by
$$[B,C]_W^{\cal D}(\Psi,\Phi):=(B^\ast \Psi, C\Phi)_{\cal W}
-(C^\ast\Psi, B\Phi)_{\cal W},\ \ \ \Psi,\Phi\in {\cal D}.$$
Assumption (A.2) is as follows.
\bi
\item[(A.2)] There exists an operator
$$T(k):\hFFF\rightarrow \hFFF,\ \ \ a. e.\  k\in M,$$
such that
\bi
\item[(1)] $D(\wh )\subset D(T(k))$ for almost every  $k\in M$, 
\item[(2)] $T(k)\Phi$, $\Phi\in D(\wh)$,  is weakly measurable in $k$, 
\item[(3)] $\displaystyle   [\i \otimes\ha (f),  \hi]_W^{D(\wh )}(\Psi,\Phi)
=\int_M\ov{f(k)} (\Psi, T(k)\Phi)_{\hFFF}d\mu(k)$ for $\Psi,\Phi\in D(\wh)$. 
\ei
\ei
Define
$$\ha _t(f):=e^{-it\wh }e^{it\wh _0}
(\i \otimes \ha (f))e^{-it\wh _0}e^{it\wh }
= e^{-it\wh }(\i \otimes \ha (e^{it \fs }f))e^{it\wh }.$$
We want to investigate $\ha_t(f)$ as $t\rightarrow \infty$ strongly.
Let $\s(S)$, $\s_{\rm p}(S)$ and $\s_{\rm ac}(S)$ be 
the spectrum, the point spectrum, and the absolutely continuous spectrum of $S$, respectively.
Assumptions (A.3)-(A.5) are as follows. 
\bi
\item[(A.3)]
The operator $S$ is purely absolutely continuous. 
\item[(A.4)]
There exists a dense subspace $\cd\subset\LM$ such that 
\bi
\item[(1)]
$\cd\subset {\cal M}_{-\han}$, and 
for any $f\in\mmm$, 
there exists a sequence $\{ f_m \}_m\subset \cd$ such that 
$\mbox{\rm s-}\!\limm f_m=f$ and $\mbox{\rm s-}\!\limm f_m/\sqrt\fs=f/\sqrt\fs$, 
\item[(2)]
for $f\in\cd$ and $\Psi\in D(\wh )$,
$$
 \int_M \ov{f(k)}
(\Psi, e^{-is(\wh -E(\wh )+\fs(k))}
T(k)\gr)_\hFFF d\mu(k) \in L^1([0,\infty),ds).$$
\ei
\item[(A.5)]
$\| T(\cdot)\gr\|_\hFFF\in\mmm$. 
\ei
\bl{zero1}
Suppose (A.3). Then for any $a\in\RR$, 
$$\mu(\{k\in M| \fs(k)=a\})=0.$$
\el
\proof
Let ${\cal N}:=\{k\in M| \fs(k)=a\}$ and suppose that 
$0< \mu({\cal N})\leq \infty $.
Since $\mu$ is a $\sigma$-finite measure, $M=\cup_{n=1}^\infty M_n$ with 
$\mu(M_n)<\infty $ for all $n$. 
Then there exists $m$ such that $\mu(M_m\cap {\cal N})>0$. 
Let $1_{M_m\cap {\cal N}}$ 
be the characteristic function on $M_m\cap {\cal N}$. Then
$1_{M_m\cap {\cal N}}\in\LM$ and 
$\fs 1_{M_m\cap {\cal N}}=
a 1_{M_m\cap {\cal N}}$, which implies that 
$a\in \s_{\rm p}(\fs)$. It contradicts with (A.3). 
Thus $\mu({\cal N})=0$. \qed
\bl{c6}
Assume (A.1)-(A.5). Then for $f\in\mmm$,
$$\int_M\|\ov{f(k)}(\wh -E(\wh )+\fs(k))\f T(k)\gr\|_\hFFF
d\mu(k)<\infty$$
and
$$(\i\otimes \ha (f))
\gr=-g\int_M\ov{f(k)} (\wh -E(\wh )+\fs(k))\f T(k)\gr d\mu(k),$$
where the integral on the right-hand side above is in the strong sense.
\el
\begin{remark}
\bi
\item[(1)] Lemma \ref{c6} is a generalization of \cite{h1, h2}. 
\item[(2)] By Lemma \ref{zero1}, $\fs(k)\not=0$ for almost every $k\in M$. Thus  
$(\wh -E(\wh )+\fs(k))\f$ is a bounded operator for almost every $k\in M$. 
\item[(3)] $T(k)\Phi$, $\Phi\in D(\wh)$,  
 is strongly measurable since $T(k)\Phi$ is weakly measurable \cite[Theorem IV.22]{rs1}. 
\ei
\end{remark}
\proof
Note that
\eqnn
&&
\int_M
\|\ov{f(k)} (\wh -E(\wh )+\fs(k))\f T(k)\gr\|_\hFFF d\mu(k)\nonumber \\
&&\label{Q}
\leq
\|f/\sqrt\fs\|_\LM\lk\int_M \|T(k)\gr\|_\hFFF^2/\fs(k)
d\mu(k)\rk^\han<\infty,
\ennn
and $\gr\in D(\i\otimes a(f))$ follows from the fact that $\gr\in
D(\i\otimes \dg(\fs))$ and Lemma \ref{arai}.
We see that for $\Psi,\Phi\in D(\wh)$ and for $f\in\cd$, 
\eqn
\frac{d}{dt}(\Psi, \ha _t(f)\Phi)_\hFFF
&=&
ig[\i\otimes \ha (e^{it\fs}f),  \hi]_W^{D(\wh )}(e^{it\wh }\Psi,
e^{it\wh }\Phi)\\
&=&
ig\int_M\ov{f(k)} e^{-it\fs(k)}(\Psi, e^{-it\wh } T(k) e^{it\wh }
\Phi)_\hFFF d\mu(k).
\enn
Then
\eqnn
&&\hspace{-1.3cm}
(\Psi, \ha_t(f)\Phi)_\hFFF\nonumber \\
&&\hspace{-1.3cm}\label{p}
=(\Psi, (\i\otimes \ha (f)) \Phi)_\hFFF+
ig\int_0^t ds \lk
\int_M\ov{f(k)} e^{-is\fs(k)}(\Psi, e^{-is\wh } T(k) e^{is\wh }
\Phi)_\hFFF d\mu(k)\rk.
\ennn
Since,  by (A.2),
$$\mbox{\rm s-}\!\limt  a_t(f)\gr=0,\ \ \ f\in\LM,$$
it follows from \kak{p} and (A.4) that 
\eqn
&&(\Psi, (\i\otimes \ha (f)) \gr)_\hFFF=-ig
\int_0^\infty ds \lk
\int_M\ov{f(k)} (\Psi, e^{-is(\wh -E(\wh )+\fs(k))} T(k)
\gr)_\hFFF d\mu(k)\rk.
\enn
We have
\eqn
&&(\Psi, (\i\otimes \ha (f)) \gr)_\hFFF\\
&& =-ig \lim_{\epsilon\rightarrow 0}
\int_0^\infty ds e^{-\epsilon s} \lk
\int_M\ov{f(k)} (\Psi, e^{-is(\wh -E(\wh )+\fs(k))} T(k) \gr)_\hFFF
d\mu(k)\rk\\
&&=-g  \int_M(\Psi, \ov{f(k)}(\wh -E(\wh )+\fs(k))\f T(k)\gr)_\hFFF
d\mu(k).
\enn
Here on the first equality we used 
the Lebesgue dominated convergence theorem and (A.4)-(2), 
on the second equality, Fubini's theorem and (A.5).
Hence by \kak{Q}, for $f\in\cd$, 
$$
(\Psi, (\i\otimes \ha (f))\gr)_\hFFF=
(\Psi, -g  \int_M\ov{f(k)}(\wh -E(\wh )+\fs(k))\f T(k)\gr
d\mu(k))_\hFFF$$
and then 
\eq{212}
(\i\otimes \ha (f))\gr= 
 -g  \int_M\ov{f(k)}(\wh -E(\wh )+\fs(k))\f T(k)\gr.
\en 
We can take  a sequence $\{ f_m\} \subset \cd$ such that 
$\mbox{\rm s-}\!\limm f_m=f$ and 
$\mbox{\rm s-}\!\limm f_m/\sqrt\fs=f/\sqrt\fs$ 
for $f\in\mmm$. 
We see that 
\eq{213}
\mbox{\rm s-}\!\limm (\i\otimes\ha(f_m))\gr=(\i\otimes\ha(f))\gr
\en 
by 
$$\|\ha(f)\Psi\|_{\hFFF}\leq \|\fs^{-\han}f\|_\LM \|\dg(\fs)^\han\Psi\|_\hFFF,\ \ \ \Psi\in D(\dg(\fs)^\han).$$
Moreover 
from 
\eqn
&& \|\int_M(\ov{f_m}(k)-\ov{f}(k)) (\wh -E(\wh )+\fs(k))\f T(k)\gr d\mu(k)\|\\
&&\leq 
\| (\ov{f_m}(k)-\ov{f}(k))/\sqrt{\fs}\|_\LM 
\lk 
\int_M \| T(k)\gr\|_\hFFF^2 /\fs(k) d\mu(k) \rk^\han,
\enn
it follows that 
\eqnn
&&
\mbox{\rm s-}\!\limm 
\int_M\ov{f_m(k)} (\wh -E(\wh )+\fs(k))\f T(k)\gr d\mu(k)\nnn \\
&&\label{214}
=
\int_M\ov{f(k)} (\wh -E(\wh )+\fs(k))\f T(k)\gr d\mu(k).
\ennn 
By \kak{213} and \kak{214}, \kak{212} can be extended for $f\in\mmm$. 
Thus the lemma  is proven.
\qed
We want to find  a necessary and sufficient conditions for
$\gr\in D(\i \otimes \dg (G)^\han)$ with
a nonnegative multiplication operator $G$ on  $\LM$.
We define $\kappa_G(k)\in\hFFF$ by
$$\kappa_G(k):=\sqrt{G(k)}(\wh -E(\wh )+\fs(k))\f T(k)\gr,\ \ \ a.e.\ k\in
M,$$
and $T_G:\LM\rightarrow \hFFF$ by
$$T_Gf :=\int _Mf(k) \kappa_G(k) d\mu(k),\ \ \ f\in\LM.$$
Then by Lemma \ref{c6}, for $f\in\mmm$, 
$$(\i\otimes \ha (\sqrt G f))\gr=-g T_G \bar f.$$
\bl{c7}
(1) $T_G$ is  a Hilbert-Schmidt operator if and only if
\eq{c30}
\int_M\|\kappa_G(k)\|_\hFFF^2 d\mu(k)<\infty.
\en
(2) Suppose that $T_G$ is a Hilbert-Schmidt operator. Then
$${\rm Tr} (T^\ast_G T_G)=\int_M\|\kappa_G(k)\|_\hFFF^2 d\mu(k).$$
\el
\proof
It is enough to show (1) and (2) for adjoint operator $T^\ast_G:\hFFF\rightarrow \LM$
instead of $T_G$.
$T^\ast_G$ 
is refered to as a Carleman operator with kernel $\kappa_G$, i.e.,
$$T^\ast_G\Phi(\cdot)=(\kappa_G(\cdot),\Phi)_\hFFF,\ \ \ \Phi\in\hFFF.$$
Then  it is established in e.g., \cite[p.141]{wei} that
$T^\ast_G$ is a
Hilbert-Schmidt operator if and only if \kak{c30} holds,
and
$${\rm Tr} (T_G T^\ast_G)=\int_M\|\kappa_G(k)\|_\hFFF^2 d\mu(k).$$
Hence the lemma is proven. \qed
\bl{1}
Let $\T$ be a bounded  operator on $\www$, 
and $\{e_m\}_{m=1}^\infty$ a  complete orthonormal system in $\www$ 
such that $e_m\in \mmm$. 
Then (1) and (2) are equivalent.
\bi
\item[(1)] $\Psi\in D(\dg (\T^\ast \T )^\han)$.
\item[(2)] $\displaystyle \Psi\in \bigcap_{m=1}^\infty D(a(\T^\ast e_m))$
and
$\displaystyle \sum_{m=1}^\infty\| a( \T^\ast
e_m)\Psi\|_{\fff(\kkk)}^2<\infty$.
\ei
Furthermore suppose that (1) or (2) is fulfilled. Then
$$\|\dg(\T^\ast \T)^\han \Psi\|_{\fff(\kkk)}^2=
\sum_{m=1}^\infty\| a( \T^\ast e_m)\Psi\|_{\fff(\kkk)}^2.$$
\el
\proof See Appendix. 
\qed
\bt{c8}
Assume (A.1)-(A.5). Let $G$ be a nonnegative multiplication operator on
$\LM$.
Then (1) and (2) are equivalent.
\bi
\item[(1)] $\gr\in D(\i\otimes \dg(G)^\han)$.
\item[(2)] $\int_M G(k)\|(\wh -E(\wh )+\fs(k))\f
T(k)\gr\|_\hFFF^2d\mu(k)<\infty$.
\ei
Furthermore suppose  that (1) or (2) holds. Then it follows that
\eqnn
\label{c10}
\|(\i\otimes \dg (G)^\han)\gr\|_\hFFF^2=
g^2 \int_M G(k)\|(\wh -E(\wh )+\fs(k))\f T(k)\gr\|_\hFFF^2d\mu(k).
\ennn
\et
\proof
We divide a proof into two steps.

\noindent 
(Step I) The case where  $G$ is bounded.

\noindent 
Let  $\{e_m\}_{m=1}^\infty$ be a orthonormal system of $\kkk$ such that $e_m\in\mmm$, 
$m\geq 1$. 
By Lemma \ref{1},
  $\gr\in D(\i\otimes \dg(G)^\han)$ is equivalent to
\eq{c9}
\sum_{m=1}^\infty \|(\i\otimes \ha (\sqrt G e_m))\gr\|^2_\hFFF<\infty,
\en
and, if \kak{c9} holds,
the left-hand side of \kak{c9} is identical with
$\|(\i\otimes \dg (G)^\han)\gr\|^2_\hFFF$.
By the definition of $T_G$,  \kak{c9} can be rewritten as
$$g^2 \sum_{m=1}^\infty \|T_G \ov{e_m}\|_\hFFF^2<\infty.$$
Then $T_G$ is a Hilbert-Schmidt operator. Hence by Lemma \ref{c7},  (1) is
equivalent to
\eq{c5}
\int_M\|\kappa_G(k)\|^2_\hFFF d\mu(k)<\infty.
\en
Namely (1) is equivalent to (2).

\noindent 
(Step II) The case where  $G$ is unbounded.

\noindent 
Let $\gr=\{\gr^{(n)}\}_{n=0}^\infty$ and
$$G_\Lambda(k):=\lkk\begin{array}{ll} G(k),& G(k)<\Lambda,\\
\Lambda, &G(k)\geq \Lambda.\end{array}\right.$$
{\it Proof of } $(1)\Longrightarrow (2)$. Note that 
\eqn
&&\|(\i\otimes \dg_n(G_\Lambda)^\han) \gr^{(n)}\|^2_{\hFFF_n} \\
&&=\int_{M^n}
\lk \sum_{j=1}^n G_\Lambda(k_j) \rk \| \gr^{(n)}(k_1,\cdots,k_n)\|_\hhh^2
d\mu(k_1)\cdots d\mu(k_n).
\enn
Hence we see that
$$\|(\i\otimes \dg(G_\Lambda)^\han )\gr\|_\hFFF^2=
\sum_{n=0}^\infty \|(\i \otimes \dg_n(G_\Lambda)^\han)
\gr\nn\|^2_{\hFFF_n}$$
is monotonously  increasing in $\Lambda$.
Then the monotone convergence theorem yields that
\eqn
&&
\lim_{\Lambda\rightarrow\infty}
\|(\i\otimes \dg(G_\Lambda)^\han )\gr\|_\hFFF^2
=
\lim_{\Lambda\rightarrow\infty}\sum_{n=0}^\infty
\|(\i\otimes \dg_n(G_\Lambda)^\han )\gr^{(n)}\|_{\hFFF_n}^2\\
&&
=\sum_{n=0}^\infty \lim_{\Lambda\rightarrow\infty}
\|(\i\otimes\dg_n(G_\Lambda)^\han )\gr^{(n)}\|_{\hFFF_n}^2
=\|(\i\otimes \dg(G)^\han )\gr\|_\hFFF^2.
\enn
Since $\gr\in D(\i\otimes \dg(G_\Lambda)^\han) $,
we have by (Step I),
$$
\|(\i\otimes \dg (G_\Lambda )^\han)\gr\|_\hFFF^2=
\int_M G_\Lambda (k)\|(\wh -E(\wh )+\fs(k))\f T(k)\gr\|_\hFFF^2d\mu(k).
$$
Take $\Lambda\rightarrow \infty$ on the both sides above.
By the monotone convergence theorem again, we obtain that
$$
\infty>\|(\i\otimes \dg (G)^\han)\gr\|_\hFFF^2
=\int_M G(k)\|(\wh -E(\wh )+\fs(k))\f T(k)\gr\|_\hFFF^2d\mu(k).
$$ 
Thus (2) follows.

{\it Proof of } $(1)\Longleftarrow (2)$.
 From (2) it follows that
$$
\infty>
\int_M G_\Lambda (k)\|(\wh -E(\wh )+\fs(k))\f T(k)\gr\|_\hFFF^2d\mu(k)
=\sum_{n=0}^\infty \|(\i\otimes \dg_n
(G_\Lambda )^\han)\gr^{(n)}\|_{\hFFF_n}^2.
$$
Then by the monotone convergence theorem, as $\Lambda\rightarrow \infty$ on
the both sides above we obtain that
\eqn
&&
\infty>\int_M G(k)\|(\wh -E(\wh )+\fs(k))\f T(k)\gr\|_\hFFF^2d\mu(k)\\
&&
= \sum_{n=0}^\infty \|(\i\otimes \dg_n (G)^\han) \gr^{(n)}\|_{\hFFF_n}^2
=\|(\i\otimes \dg(G)^\han) \gr\|_\hFFF^2.
\enn
Thus (1) follows.

Finally \kak{c10} follows from the fact that 
\eqn
&& \|(\i\otimes  \dg(G)^\han)\gr\|_\hFFF^2=
g^2\sum_{m=1}^\infty\|T_G\ov{e_m}\|_\hFFF^2=
g^2{\rm Tr}(T^\ast_G T_G)\\
&&
=
g^2 \int_M G(k)\|(\wh -E(\wh )+\fs(k))\f T(k)\gr\|_\hFFF^2d\mu(k).
\enn
Thus the proof is complete. 
\qed

\bc{ko2}
Let $G$ be a nonnnegative multiplication operator on $\LM$.
In addition to (A.1)-(A.5), suppose that 
\eq{irr}
{\sqrt G}\|T(\cdot)\gr\|_\hFFF/{\fs}\in \LM.
\en
Then
$\gr\in D(\i\otimes\dg (G)^\han)$.
In particular suppose that 
\eq{irrr}
\|T(\cdot)\gr\|_\hFFF/{\fs}\in \LM.
\en
Then 
$\gr\in D(\i\otimes N^\han)$.
\ec
\begin{remark}
Condition \kak{irr} is a generalization  of an IR regularity condition in \cite{ah2}.
\end{remark}
\proof
Since
$$
\int_M G(k)\|(\wh -E(\wh )+\fs(k))\f T(k)\gr\|_\hFFF^2d\mu(k)\leq
\int_M \frac{G(k)}{\fs(k)^2}\|T(k)\gr\|_\hFFF^2 d\mu(k)<\infty,$$
Theorem \ref{c8} yields the corollary. 
\qed

\section{Absence of ground states}
Let $\pg$ be the projection onto
the one-dimensional subspace $\{z\gr | z\in\CC\}$ of $\hFFF$.
\bt{d1}
Assume (A.1)-(A.5).
Let $G$ be a nonnegative multiplication operator on $\LM$.
Then $\wh $ has no ground state $\gr$ in $D(\i \otimes \dg (G)^\han)$ such
that
\eq{d2}
{\sqrt G}
(\gr, T(\cdot)\gr)_\hFFF/{\fs}\not \in \LM.
\en
\et
\begin{remark}
Condition \kak{d2} is a generalization of an IR singularity condition in \cite{ah2}.
\end{remark}
\proof
Suppose that there exists a ground state $\gr$ such as  in \kak{d2}.
We have
\eqn
&&\int_M G(k)\|(\wh -E(\wh )+\fs(k))\f T(k) \gr\|_\hFFF^2 d\mu(k)\\
&&\geq
\int_M G(k)\|(\wh -E(\wh )+\fs(k))\f \pg T(k) \gr\|_\hFFF^2 d\mu(k).
\enn
Since
\eqn
&&
\|(\wh -E(\wh )+\fs(k))\f \pg T(k) \gr\|_\hFFF^2 \\
&&=\frac{1}{\fs(k)^2}\|\pg T(k) \gr\|_\hFFF^2
=\frac{1}{\fs(k)^2}|(\gr, T(k)\gr)_\hFFF|^2,
\enn
it follows that
$$
\int_M G(k)\|(\wh -E(\wh )+\fs(k))\f T(k) \gr\|_\hFFF^2 d\mu(k)
\geq
\int_M\lk \frac{\sqrt{G(k)} }{\fs(k)}|(\gr, T(k)\gr)_\hFFF|\rk ^2d\mu(k).$$
Since the right-hand side above diverges  by \kak{d2}, Lemma \ref{c7}
yields that
$$\gr\not\in D(\i\otimes \dg(G)^\han).$$
Thus the desired result is obtained.
\qed
Setting $G=1$ in Theorem \ref{d1}, we have a corollary. 
\bc{ko222} 
Assume (A.1)-(A.5).
Then $\wh $ has no ground state $\gr$ in $D(\i \otimes N^\han)$ such
that
\eq{d2d}
\frac{1}{\fs} (\gr, T(\cdot)\gr)_\hFFF\not \in \LM.
\en
\ec

\section{Higher order regularities of ground states}
Through this section we fix a natural number $n$ 
and 
assume (A.1)-(A.5).
We shall consider an asymptotic field for
$\i\otimes \ha  (f_1)\cdots \ha (f_n)$.
Namely we investigate
$$\mbox{\rm s-}\!\limt e^{-it\wh } (\i\otimes
\ha(e^{it\fs} f_1)\cdots \ha(e^{it\fs} f_n) ) e^{it\wh }.$$

We introduce assumptions.
\bi
\item[(H.1)] $T(k)$ satisfies that 
$$D(\wh)\subset D(T(k)(\i\otimes\ha(f)))\cap D((\i\otimes \ha(f))T(k))\ \ \ a.e. k\in M,$$
and 
\eq{ko12}
[T(k), \i\otimes \ha(f)]\Psi=0,\ \ \ \ \Psi\in D(\wh),\ \ \ a.e.\ k\in M.\en
\item[(H.2)] The operator 
$$( \i\otimes\dg(\fs)^{(n+1)/2}) (\wh +z)^{-m}$$ is a bounded
operator for some  $m$ and   $z\in \rho(H)\cap\RR$, where $\rho(H)$ 
denotes the resolvent of $H$.
\ei
\bl{arai3}
Suppose (H.2). Then
for $f_j\in\mm$, $j=1,...,n$,
\eq{ko10}
\gr\in D(\i\otimes \ha(f_1)\cdots \ha(f_n))
\en
and 
\eq{ko100}
(\i\otimes \ha(f_1)\cdots \ha(f_m))\gr\in D(\wh),\ \ \ m=1,...,n-1.
\en 
\el
\proof
From (H.2) it follows that
$$ 
\|(\i\otimes\dg(\fs)) ^{(n+1)/2}\gr\|_\hFFF
\leq 
\|(\i\otimes\dg(\fs)) ^{(n+1)/2} (\wh +z)^{-m}\|_{\hFFF\rightarrow\hFFF}
\left|E(\wh )+z\right|^m \|\gr\|_\hFFF.
$$ 
Hence
\eq{ko9}
\gr \in D(\i\otimes\dg(\fs)^{(n+1)/2})
\en  follows, from which
we can see \kak{ko10} by Lemma \ref{arai}.
From Lemma \ref{arai} it follows that 
$$(\i\otimes \ha(f_1)\cdots \ha(f_n))\gr\in D(\i\otimes \dg(\fs)).$$
Together with 
$$(\i\otimes \ha(f_1)\cdots \ha(f_n))\gr\in D(A\otimes\i),$$
we obtain that 
$$ (\i\otimes \ha(f_1)\cdots \ha(f_n))\gr\in D(A\otimes\i)\cap D(\i\otimes\dg(\fs))=D(\wh).$$
Then \kak{ko100} is proven. 
\qed
${\cal P}_n$ denotes the set of all the permutations of degree $n$ and we set for
$\s\in {\cal P}_n$,
\eqn
&& R_i^\s:=
\lk \wh -E(\wh )+\sum_{j=i}^n \fs(k_{\sigma(j)})\rk  \f ,\\
&& \tilde  R_i^\s:=
\lk \wh -E(\wh )+ \fs(k_{\sigma(i)})\rk  \f .
\enn
\begin{remark}
Let $\otimes^m\mu$ denote the product measure on $M^m$. 
Then 
$$(\otimes^m \mu) 
(\{(k_1,...,k_m) \in M^m | \fs(k_1)+\cdots + \fs(k_m) =0\})=0$$
follows. This implies that $R_i^\s$ and $\tilde  R _i^\s$ are bounded operators on $\hFFF$  
for almost every 
$(k_{\s(i)},...,k_{\s(n)})\in M^{n-i+1}$.
\end{remark}
We set 
$$ \oh:=\wh-E(\wh).$$
Assumption (H.3) and (H.4) are  as follows. 
\bi
\item[(H.3)]
There exist dense subspaces  $\ccd\subset\LM$ 
and ${\cal E}\subset\hFFF$ such that 
\bi
\item[(1)] $\ccd\subset \mm$, and for $f\in\mm$, 
there exists a sequence $\{f_m\}_m\subset \ccd$ such that
$\mbox{\rm s-}\!\limm f_m/\sqrt{\fs^k}=f/\sqrt{\fs^k}$ for $0\leq k\leq n$,
\item[(2)] 
$\Psi \subset \bigcap_{m=1}^n D(\q m)$
for almost every $(k_1,...,k_m)\in M^m$ and for $\Psi\in{\cal E}$, 
\item[(3)] 
for  arbitrary $f_j\in\ccd$, $j=1,...,n$, and $\Psi\in{\cal E}$, 
\eqn 
&&\hspace{-1cm} \int_Md\mu(k_m) e^{-i T_m  \fs(k_{m})} f_m(k_m) \\
&&
 (\q m\Psi,  (\i\otimes \ha(e^{iT_m \fs}f_{m+1})
\cdots 
 \ha(e^{iT_m  \fs}f_n) ) \gr)_\hFFF
\enn is in $L^1([0,\infty);dt_m)$  
for $m=1,2,...,n-1$, 
where 
$ T_m:=t_1+\cdots+t_m$,
\item[(4)] 
for  arbitrary $f_j\in\ccd$, $j=1,...,n$, and $\Psi\in{\cal E}$, 
\eqn 
&&\hspace{-2cm}
\int_M d\mu(k_{\s(m)}) \cdots \int_M d\mu(k_{\s(n)}) 
\prod_{j=m}^n f_{\s(k)}(\ks) \\
&&
(\qm m \Psi, \\
&& 
\hspace{4cm} R_{m+1}^\s T(k_{\s(m+1)}) \cdots 
R_{n}^\s T(k_{\s(n)})\gr)_\hFFF
\enn is in $L^1([0,\infty);dt_m)$  
for $m=1,2,...,n-1$. 
\ei
\item [(H.4)]
The closure of $(\wh-E(\wh)+\fs(k))\f T(k)$, 
$[(\wh-E(\wh)+\fs(k))\f T(k)\ov]$,
is a bounded opeator and 
$$\|[(\wh-E(\wh)+\fs(\cdot ))\f T(\cdot )\ov]
\|_{\hFFF\rightarrow\hFFF}\in{\cal M}_\han.$$
\ei
\bl{k10}
Assume (H.1)-(H.4). Then for
$f_\ell\in\mm$, $\ell=1,...,n$,
\eq {ko4}
\int_{M^n} \prod_{j=1}^n d\mu(k_j)
\lk\prod_{j=1}^n |{f_j(k_j)}| \rk
\left\|
R_1^\s T(k_{\sigma(1)})
\cdots
R_n^\s T(k_{\sigma(n)})
\gr\right\|_\hFFF<\infty
\en
and
\eqnn
&&( \ii \ha(f_1) \cdots \ha(f_n) ) \gr\nonumber \\
&&
\label{ko13}
= (-g)^n \sum_{\sigma\in{\cal P}_n}
\int_{M^n} \prod_{j=1}^n d\mu(k_j)
\lk
\prod_{j=1}^n \ov{f_j(k_j)} \rk
R_1^\s T(k_{\sigma(1)})
\cdots
R_n^\s T(k_{\sigma(n)})
\gr.
\ennn
\el
\proof
Note that 
$$\|R_j^\s\|_{\hFFF\rightarrow \hFFF}
\leq \|\tilde {R}_j^\s\|_{\hFFF\rightarrow \hFFF},\ \ \ j=1,...,n.$$
From (H.4) it follows that 
\eq{49}
f(\ks)\|[{\tilde R_j^\s T(\ks)}\ov]\|_{\hFFF\rightarrow\hFFF}\in L^1(M)
\en 
for $f\in \mmm$ and $\s\in{\cal P}_n$.
We have,  by \kak{49}, 
\eqnn
&&\int_{M^n} \prod_{j=1}^n d\mu(k_j)
\lk
\prod_{j=1}^n |{f_j(k_j)}| \rk
\left\|
R_1^\s T(k_{\sigma(1)})
\cdots
R_n^\s T(k_{\sigma(n)})
\gr\right\|_\hFFF\nonumber \\
&&
\label{fum}
\leq \|\gr\|_\hFFF
\prod_{j=1}^n
\int_M | \ff j(k_{\s(j)})| \| 
[
\tilde 
R_j^\s T(\ks)\ov] \|_{\hFFF\rightarrow\hFFF}
d\mu(\ks) <\infty.
\ennn
Then \kak{ko4} follows.
Lemma \ref{arai} yields that
the left-hand side of \kak{ko13} is well defined.
Let $f_j\in\ccd$, $j=1,...,n$. 
Note that by \kak{ko100}, 
$$(\i\otimes \ha(f_1)\cdots \ha(f_m))\gr\in 
D(\wh),\ \ \ \ m=1,...,n-1.$$
From this and (H.1) it follows that 
\eq{haha}
[T(k), \i\otimes \ha(g)](\i\otimes \ha(f_1)\cdots \ha(f_m))\gr=0.
\en 
By \kak{haha} we see that  for $\Psi\in{\cal E}$ and $f_j\in\ccd$, $j=1,...,n$, 
\eqn
&& \frac{d}{dt}(\Psi,
e^{-it\wh } (\ii \ha(e^{it\fs} f_1)\cdots \ha(e^{it\fs} f_n) )e^{it\wh }
\gr)_\hFFF\\
&&
=
 ig\sum_{j=1}^n \int _M d\mu(k_j) e^{-it\fs(k_j)}f_j(k_j)\\ 
&&  (e^{it\oh }\Psi,
( \i\otimes  \ha(e^{it\fs}f_1)\cdots {\ha(e^{it\fs}f_{j-1})}) T(k_j) 
(\i\otimes {\ha(e^{it\fs}f_{j+1})}
\cdots \ha(e^{it\fs}f_n)) \gr)_\hFFF\\
&&
= ig\sum_{j=1}^n \int _M d\mu(k_j) e^{-it\fs(k_j)}f_j(k_j)\\ 
&&\hspace{1cm}  (e^{it\oh }\Psi,
 T(k_j)( \i\otimes  \ha(e^{it\fs}f_1)\cdots \sl{\ha(e^{it\fs}f_j)}
\cdots \ha(e^{it\fs}f_n))\gr)_\hFFF.
\enn
Then by (H.3)-(3) and the fact that  
$$\mbox{\rm s-}\!\lim_{t\rightarrow \infty} 
(\Psi, e^{-it\wh} 
(\i\otimes\ha(e^{it\fs}f_1)\cdots \ha(e^{it\fs}f_n))
e^{itE(\wh)} \gr)=0,$$
we have 
\eqnn
&&(\Psi, (\i\otimes \ha(f_1)\cdots\ha(f_n))\gr)_\hFFF\nonumber \\
&&=
 -ig\sum_{j=1}^n \int_0^\infty dt \int _M d\mu(k_j)
e^{-it\fs(k_j)}\ov{f_j(k_j)}
\nonumber \\
&&
\label{ko5}
\hspace{2cm} 
(T(k_j)^\ast e^{it\oh}    \Psi,   (\i\otimes  \ha(e^{it\fs}f_1)\cdots
\sl {\ha(e^{it\fs}f_j)}
\cdots a(e^{it\fs}f_n)) \gr)_\hFFF.
\ennn
Using \kak{ko5} again,  we have by  (H.3)-(3), 
\eqn
&&(T(k_j)^\ast   e^{it\oh} \Psi, 
(\i\otimes \ha(e^{it\fs}f_1)\cdots \sl {\ha(e^{it\fs}f_j)}
\cdots \ha(e^{it\fs}f_n)) \gr)_\hFFF\\
&&=
 -ig\sum_{j'=1, j'\not=j}^n
\int_0^\infty dt' \int _M d\mu(k_{j'}) e^{-i(t+t')\fs(k_{j'})}\ov{f_{j'}(k_{j'})}
(T(k_{j'})^\ast e^{it'\oh} T(k_j)^\ast e^{it\oh} \Psi, 
\\
&&\hspace{1cm}
(\i\otimes
\ha(e^{i(t+t')\fs}f_1)\cdots \sl {\ha(e^{i(t+t')\fs}f_j)} \cdots
\sl {\ha(e^{i(t+t')\fs}f_{j'})}
\cdots \ha(e^{i(t+t')\fs}f_n)) \gr)_\hFFF.
\enn
We  can inductively obtain that 
\eqn
&&(\Psi, (\i\otimes \ha(f_1)\cdots\ha(f_n))\gr)_\hFFF\\
&&=(-ig)^n\sum_{\s\in{\cal P}_n}
\int_0^\infty dt_1\int_M d\mu(k_{\s(1)})\cdots 
\int_0^\infty dt_n\int_M d\mu(k_{\s(n)})\lk \prod_{j=1}^n\ov{f_{\s(j)}(\ks)} \rk  \\
&& \hspace{3cm} 
e^{-it_1\sum_{j=1}^n \fs(\ks)} e^{-it_2 \sum_{j=2}^n \fs(\ks)} \cdots e^{-it_n \fs(k_{\s(n)})} \\
&&\hspace{6cm}
( \q n\Psi, \gr)_\hFFF\\
&&=(-ig)^n\sum_{\s\in{\cal P}_n}
\int_0^\infty dt_1\int_M d\mu(k_{\s(1)})\cdots 
\int_0^\infty dt_n\int_M d\mu(k_{\s(n)})\lk \prod_{j=1}^n\ov{f_{\s(j)}(\ks)} \rk  \\
&& \hspace{2cm}  (\qm n\Psi, \gr)_\hFFF.
\enn
Since by (H.3)-(4), 
\eqn 
&& 
\hspace{-1cm}\int_M d\mu(k_{\s(n)})
\prod_{j=1}^n\ov{f_{\s(j)}(\ks)}\\
&&
\hspace{1cm}
(\qm n\Psi, \gr)_\hFFF
\enn
is in $ L^1([0,\infty),dt_n)$, 
we have 
\eqn
&&
\int_0^\infty dt_n\int_M d\mu(k_{\s(n)})
\lk\prod_{j=1}^n\ov{f_{\s(j)}(\ks)}\rk\\
&&
\hspace{1cm} 
(\qm {n-1} \Psi, \\
&&
\hspace{8cm} 
e^{-it_n (\oh+\fs(k_{\s(n)}))}   T(k_{\s(n)})\gr)_\hFFF\\
&&=
\lim_{\epsilon\rightarrow 0} 
\int_0^\infty dt_ne^{-\epsilon t_n} 
\int_M d\mu(k_{\s(n)})
\lk\prod_{j=1}^n\ov{f_{\s(j)}(\ks)}\rk\\
&&
\hspace{1cm}  
(\qm {n-1} \Psi, \\
&& \hspace{8cm} 
e^{-it_n (\oh+\fs(k_{\s(n)}))}   T(k_{\s(n)})\gr)_\hFFF\\
&&
=(-i) \int_M d \mu(k_{\s(n)})\lk\prod_{j=1}^n\ov{f_{\s(j)}(\ks)}\rk\\
&&
\hspace{1cm}  (\qm {n-1}\Psi, \\
&&\hspace{8cm} 
R_n^\s
T(k_{\s(n)})\gr)_{\hFFF}.
\enn
Here we used the Lebesgue dominated convergence theorem 
and Fubini's lemma. 
Similarly using  (H.3)-(4) we can obtain that 
\eqn
&&
\int_0^\infty dt_{n-1}\int_M d\mu(k_{\s(n-1)})
\int_0^\infty dt_n\int_M d\mu(k_{\s(n)})
\lk \prod_{j=1}^n\ov{\ff j (\ks)} \rk \\
&&
\hspace{1cm}
(\qm n  \Psi, \gr)_{\hFFF}\\
&&=
(-i) \int_0^\infty dt_{n-1}\int_M d\mu(k_{\s(n-1)})
\int_M d\mu(k_{\s(n)})
\lk\prod_{j=1}^n\ov{\ff j (\ks)}\rk \\
&&
\hspace{1cm}(\qm {n-1}\Psi, \\ 
&&\hspace{8cm} R_n^\s  T(k_{\s(n)})\gr)_{\hFFF}\\
&&=(-i)^2 \int_M d\mu(k_{\s(n-1)})
\int_M d\mu(k_{\s(n)})
\lk\prod_{j=1}^n\ov{\ff j (\ks)}\rk \\
&&
\hspace{1cm}(\qm {n-2}\Psi, \\
&& \hspace{6cm} R_{n-1}^\s  T(k_{\s(n-1)}) 
R_n^\s  T(k_{\s(n)})\gr)_{\hFFF}.
\enn
Thus inductively we can see that 
\eqn
&&(\Psi, 
(\ii \ha(f_1) \cdots \ha(f_n) )\gr)_{\hFFF} \\
&&
= (-g)^n \sum_{\sigma\in{\cal P}_n}
\int_{M^n} \prod_{j=1}^n d\mu(k_{\sigma(j)})
\prod_{j=1}^n \ov{\ff j(k_{\sigma(j)})} 
(\Psi, 
R_1^\s T(k_{\sigma(1)})
\cdots
R_n^\s T(k_{\sigma(n)})
\gr)_\hFFF.
\enn
By \kak{fum} and the fact that ${\cal E}$ is dense, we obtain that  for $f_j\in\ccd$, 
$j=1,...,n$, 
\eqnn
&&(\ii \ha(f_1) \cdots \ha(f_n) )\gr\nnn \\
&&
= (-g)^n \sum_{\sigma\in{\cal P}_n}
\int_{M^n} \prod_{j=1}^n d\mu(k_{\sigma(j)})
\lk
\prod_{j=1}^n \ov{\ff j(k_{\sigma(j)})} \rk
R_1^\s T(k_{\sigma(1)})
\cdots
R_n^\s T(k_{\sigma(n)})
\gr\nnn \\
&&
\label{410}
= (-g)^n \sum_{\sigma\in{\cal P}_n}
\int_{M^n} \prod_{j=1}^n d\mu(k_j) 
\lk
\prod_{j=1}^n \ov{f_j(k_j)} \rk 
R_1^\s T(k_{\sigma(1)})
\cdots
R_n^\s T(k_{\sigma(n)})
\gr.
\ennn
Let $f_j\in\mm$, $j=1,...,n$. 
Take sequences $\{f_{jm}\}_m$, $j=1,...,n$, in $\ccd$  such that 
$\mbox{\rm s-}\!\limm f_{jm}/\sqrt{\fs^k}=f_j/\sqrt{\fs^k}$, $j=1,...,n$, 
 for $0\leq k\leq n$. 
Then 
\eq{411}
\mbox{\rm s-}\!\lim_{m\rightarrow\infty} 
(\ii \ha(f_{1m}) \cdots \ha(f_{nm}) )\gr=
(\ii \ha(f_1) \cdots \ha(f_n) )\gr.
\en 
Moreover by \kak{fum} it follows that 
\eqnn
&&\mbox{\rm s-}\!\lim_{m\rightarrow\infty}
\int_{M^n} \prod_{j=1}^n d\mu(k_j) 
\lk
\prod_{j=1}^n \ov{f_{jm}(k_j)} \rk 
R_1^\s T(k_{\sigma(1)})
\cdots
R_n^\s T(k_{\sigma(n)})
\gr\nnn \\
&&
\label{412}
=\int_{M^n} \prod_{j=1}^n d\mu(k_j) 
\lk
\prod_{j=1}^n \ov{f_j (k_j)} \rk 
R_1^\s T(k_{\sigma(1)})
\cdots
R_n^\s T(k_{\sigma(n)})
\gr.
\ennn
Hence \kak{410} can be extended for $f_j\in\mm$, $j=1,...,n$, by \kak{411} and \kak{412}. 
Then the lemma follows.
\qed
\bl{k1}
Let $\{e_j\}_{j=1}^\infty$ be a complete orthonormal system in $\kkk$.
Then (1) and (2) are equivalent.
\bi
\item[(1)]
$\displaystyle
\Psi\in \bigcap_{i_1,...,i_n=1}^\infty D(a(e_{i_1})\cdots a (e_{i_n}))$ and
$\displaystyle
\sum_{i_1,...,i_n=1}^\infty \|a(e_{i_1})\cdots a (e_{i_n})
\Psi\|^2_{\fff(\kkk)}<\infty.$
\item[(2)] $\displaystyle  \Psi\in D(\prod_{j=1}^n (N-j+1)^\han)$.
\ei
Suppose that (1) or (2) holds. Then
$$
\sum_{i_1,...,i_n=1}^\infty \|a(e_{i_1})\cdots a
(e_{i_n})\Psi\|^2_{\fff(\kkk)}
=\| \prod_{j=1}^n (N-j+1)^\han\Psi\|^2_{\fff(\kkk)}.$$
\el
\proof
See Appendix. 
\qed
\bt{k2}
Assume (H.1)-(H.4). Then (1) and (2) are equivalent.
\bi
\item [(1)]
$\displaystyle
\int_{M^n}\prod_{j=1}^n d\mu(k_j)
\| \sum_{\sigma\in{\cal P}_n}
R_1^\s T(k_{\sigma(1)})
\cdots
R_n^\s T(k_{\sigma(n)})\gr
   \|_\hFFF^2<\infty $.
\item [(2)] $ \displaystyle  \gr\in D (\i\otimes \prod_{j=1}^n
(N-j+1)^\han)$.
\ei
Furthermore suppose that (1) or (2) holds. Then
\eqnn
&& \|(\i\otimes  \prod_{j=1}^n (N-j+1)^\han) \gr\|^2_{\hFFF} \nonumber\\
&& \label{k15}
=g^{2n}
\int_{M^n}\prod_{j=1}^n d\mu(k_j) \| \sum_{\sigma\in{\cal P}_n}
R_1^\s T(k_{\sigma(1)})
\cdots
R_n^\s T(k_{\sigma(n)})\gr
   \|_\hFFF^2.
\ennn
\et
\proof
Let us define
$$\kappa(k_1\cdots k_n):= \sum_{\sigma\in{\cal P}_n}
R_1^\s T(k_{\sigma(1)})
\cdots
R_n^\s T(k_{\sigma(n)}) \gr,\ \ \ a.e.\ (k_1,...,k_n)\in M^n,$$
and
$T:\otimes^n\LM\rightarrow \hFFF$ by
$$T(f_1\otimes \cdots \otimes f_n):=\int _{M^n}
\prod_{j=1}^n d\mu(k_j)
f(k_1) \cdots f(k_n) \kappa(k_1\cdots k_n).$$
Then
$$(\ii \ha (f_1) \cdots \ha(f_n)) \gr=(-g)^nT(\ov{f_1}\otimes
\cdots\otimes\ov{f_n})$$
for $f_j\in\mm$, $j=1,...,n$.
Adjoint operator 
$T^\ast:\hFFF\rightarrow \otimes^n\LM\cong L^2(M^n)$ is given by
$$(T^\ast\Phi)(k_1,...,k_n) =(\kappa(k_1,...,k_n), \Phi)_\hFFF$$ 
for almost ever $(k_1,...,k_n)\in M^n$. 
Then $T^\ast$ is a Carleman operator \cite{wei} with kernel $\kappa$. Thus
$T^\ast$ is a Hilbert-Schmidt operator if and only if
$$\int_{M^n} \prod_{j=1}^n d\mu(k_j) \|\kappa(k_1\cdots k_n)\|_\hFFF^2
<\infty.$$
Namely $T$ is a Hilbert-Schmidt operator if and only if
\eq{k13}
\int_{M^n} \prod_{j=1}^n d\mu(k_j) \| \sum_{\sigma\in{\cal P}_n}
R_1^\s T(k_{\sigma(1)})
\cdots
R_n^\s  T(k_{\sigma(n)})\gr
   \|_\hFFF^2 <\infty.
\en
Let $\{e_j\}_{j=1}^\infty$ be a complete orthonormal system in $\LM$ such
that
$e_j\in\mm$, $j\geq 1$. 
If $T$ is a Hilbert-Schmidt operator, then 
\eqnn
&&\hspace{-0.5cm}
\infty>
g^{2n}
\int_{M^n} \prod_{j=1}^n d\mu(k_j) \| \sum_{\sigma\in{\cal P}_n}
R_1^\s T(k_{\sigma(1)})
\cdots
R_n^\s  T(k_{\sigma(n)})\gr
   \|_\hFFF^2 \nonumber\\
&&\hspace{-0.5cm}
=g^{2n}
\int_{M^n} \prod_{j=1}^n d\mu(k_j) \|\kappa(k_1\cdots k_n)\|_\hFFF^2
=g^{2n}
{\rm Tr}(T^\ast T)
=
g^{2n}
\sum_{i_1,...,i_n} ^\infty \|T(\ov{e_{i_1}}\otimes\cdots \otimes
\ov{e_{i_n}})\|^2_\hFFF
\nonumber \\
&&\hspace{-0.5cm}\label{k16}=
\sum_{i_1,...,i_n} ^\infty \|(\ii a(e_{i_1})\cdots a(e_{i_n}))\gr\|^2_\hFFF
=\|(\i\otimes \prod_{j=1}^n(N-j+1)^\han)\gr  \|_\hFFF^2.
\ennn
Thus  \kak{k13} is equivalent to
$\displaystyle \gr\in D(\i\otimes \prod_{j=1}^n(N-j+1)^\han)$.
Moreover \kak{k15} follows from \kak{k16}. The proof is complete.
\qed
Assumption (H.5)  is as follows. 
\bi
\item[(H.5)] The closure of $(\wh-E(\wh)+\fs(k))\f T(k)$, 
$[(\wh-E(\wh)+\fs(k))\f T(k)\ov{]}$, 
is a bounded operator for almsot every $k\in M$, and 
 $$\|[(\wh-E(\wh)+\fs(\cdot))\f T(\cdot)\ov{]}\|_{\hFFF\rightarrow \hFFF}\in
{\cal M}_0.$$ 
\ei
\bt{k4} In addition to  (H.1)-(H.4),
we assume (H.5).
Then 
$$\displaystyle
\gr\in  D(\i\otimes N^{n/2}).$$ 
\et
\proof
Note that 
\eq{lo}
D(N^{n/2})= \bigcap_{\ell =1}^n D(\prod_{j=1}^\ell(N-j+1)^\han).
\en 
See e.g., \cite[Lemma 3.2]{h17}. 
We see that for all $1\leq \ell\leq n$, 
\eqn
&& \| (\i\otimes \prod_{j=1}^\ell (N-j+1)^\han)\gr\|_\hFFF^2\\
&&
=g^{2\ell} \int_{M^\ell} \prod_{j=1}^\ell  d\mu(k_j) \| \sum_{\sigma\in{\cal P}_\ell}
R_1^\s T(k_{\sigma(1)})
\cdots
R_\ell^\s T (k_{\sigma(\ell)})\gr
   \|_\hFFF^2 \\
&&\leq \ell (\ell!)^2 g^{2\ell}  \|\gr\|_\hFFF ^2 
\lk 
\int _M
\| (\wh-E(\wh)+\fs(k))\f T(k)  \|_{\hFFF\rightarrow \hFFF}^2  d\mu(k)\rk^\ell<\infty.
\enn
Then 
$$\gr\in \bigcap_{\ell =1}^n D(\prod_{j=1}^\ell(N-j+1)^\han).$$
By \kak{lo} the theorem follows.
\qed

\section{GSB models}
\subsection{Regularities  of ground states}
The Hilbert space for the GSB model is defined by
$$\FFFg:=\hhh\otimes \fff(\LR).$$
Let $a(f)$ and $\add(f)$, $f\in\LR$,  be
the annihilation operator and the creation operator of $\fff(\LR)$,
respectively.
Set
$$\phi(\la):=\frac{1}{\sqrt2}\lk \add(\la)+a(\la)\rk,\ \ \ \la\in\LR.$$
Hamiltonians of GSB models  are defined by
$$\gsb:=\gsbz+\alpha \gsbi,$$
where $\alpha$ is a coupling constant,
$$\gsbz:=A\otimes \i+\i\otimes\dg(\omega)$$
and
$$\gsbi:=\ov{\sum_{j=1}^J B_j\otimes \phi(\la_j)}.$$
Here 
$\omega$ is a nonnegative multiplication operator on $\LR$,
$B_j$, $j=1,...,J$, closed symmetric operators on $\hhh$, and 
$\ov{\jjj B_j\otimes \phi(\la_j)}$
denotes the closure of $\jjj B_j\otimes \phi(\la_j)$.
Using fundamental inequalities
\kak{2} and \kak{3}
we have 
\eq{sq}
\|\phi(\la)\Psi\|_{\fff(\LR)}\leq 
\frac{1}{\sqrt2}(\|\la/\sqrt\omega\|_\LR+2\|\la\|_\LR) 
\|(\dg(\omega)+1)^\han\Psi\|_{\fff(\LR)}
\en 
for $\Psi\in D(\dg(\omega)^\han)$. 
Assumptions are as follows.
\bi
\item[(B.1)] $A$ is self-adjoint and bounded from below.
\item[(B.2)] $\la_j,\la_j/\sqrt\omega\in\LR$, $j=1,...,J$.
\item[(B.3)] $D(\avv  ^\han)\subset \bigcap _{j=1}^J D(B_j)$, where 
$\avv :=A-E(A)$,
and there exist constants $a_j$ and $b_j$ such that
$$\| B_j f\|_\hhh
\leq a_j\| \avv ^\han f\|_\hhh +b_j\|f\|_\hhh,\ \ \ j=1,...,J,\ \ \
f\in D(\avv ^\han).$$
Moreover
$$|\alpha|\leq \lk \sum_{j=1}^Ja_j\|\la_j/\sqrt\omega\|^2_\LR\rk\f.$$
\ei
\bp{g1}
Assume (B.1)-(B.3). Then $\gsb$ is self-adjoint
and  bounded from below on
$D(\gsbz)=D(A\otimes  \i)\cap (\i\otimes \dg(\omega))$.
Moreover it is essentially self-adjoint on any core of $\gsbz$.
\ep
\proof
It is easily seen that for $\Psi\in D(\gsbz)$,
\eq{51}
\|\gsbi \Phi\|_\FFFg
\leq
\lk \sum_{j=1}^Ja_j\|\la_j/\sqrt\omega\|^2_\LR\rk\|\gsbz\Phi\|_\FFFg
+ b\|\Phi\|_\FFFg
\en with some constant $b>0$.
Then by the Kato-Rellich theorem, the proposition follows.
\qed
To formulate additional assumptions we define
$$Y:=\bigcup_{n=1}^\nu\{k=(k_1,...,k_\nu)\in \BR| k_n=0\}.$$
We assume  (B.4) and (B.5).
\bi
\item[(B.4)] $\la_j\in C^2(\BR\setminus Y)$, $j=1,...,J$.
\item[(B.5)] 
\bi
\item[(1)]
$\omega\in C^3(\BR\setminus Y)$ and $\partial\omega(k)
/\partial k_n\not=0$ on $\BR\setminus Y$, $n=1,...,\nu$,
\item[(2)]
$\omega$ is purely absolutely continuous. 
\ei
\ei

\br{g2}
In \cite{ah1} the existence of ground states of $\gsb$
is proven under some conditions on $\omega$ for $\alpha$ with $|\alpha|$
sufficiently small.
Moreover for massive cases
the multiplicity of ground states is studied  in \cite{ah1}.
For massless cases,
an upper bound of the multiplicity of ground states is shown in \cite{hi27}.
\er
We see that,  for $\Psi,\Phi\in D(\gsbz)$,
\eq{52}
[\i\otimes a(f), \gsbi]_W^{D(\gsbz)}(\Psi, \Phi)=
\int_\BR \ov{f(k)}(\Psi, \tgsb(k)\Phi)_\FFFg dk,
\en 
where
$$\tgsb(k):=\lk \jjj \la_j(k) B_j\rk \otimes \i.$$
\bt{g3}
Assume that (B.1)-(B.5).
Let $G$ be a nonnegative multiplication operator on $\LR$.
Then
$$\gr\in D(\i\otimes\dg(G)^\han)$$ if and only if
$$\int_\BR G(k)\|(\gsb-E(\gsb)+\omega(k))\f\tgsb(k)\gr\|^2 _{\FFFg}
dk<\infty.$$
Furthermore suppose $\gr\in D(\i\otimes\dg(G)^\han)$. Then
$$\|(\i\otimes\dg(G)^\han)\gr\|^2_{\FFFg}
=\int_\BR G(k)\|(\gsb-E(\gsb)+\omega(k))\f\tgsb(k)\gr\|^2 _{\FFFg} dk.$$
\et
\proof
Under the identifications:
\eq{d5}
\wh =\gsb,\ \ \  \hi=\gsbi,\ \ \ \fs(k)=\omega(k),\ \ \
T(k)=\tgsb(k), \ \ \cd=C_0^2(\BR\setminus Y),
\en
it is enough to check (A.1)-(A.5) by Theorem \ref{c8}. 
(A.1), (A.2) and  (A.3)  follow from \kak{51},  \kak{52} and  (B.5)-(2), 
respectively. 
Since 
$$\|\tgsb(k)\gr\|_{\FFFg}\leq\jjj\la_j(k)\|(B_j\otimes\i)\gr\|_{\FFFg},$$
it is seen that
$$
\| \tgsb(\cdot)\gr\|_{\FFFg}, \| \tgsb(\cdot)\gr\|_{\FFFg}/\sqrt\omega \in \LR
$$
by (B.2). Thus (A.5) follows. 
We shall check (A.4). 
(A.4)-(1) is trivial. 
It is seen that for $k\in\BR\setminus Y$,
\eq{ki}
e^{-is\omega(k)}=-\frac{1}{s^2}
\lk\frac{\partial\omega(k)}{\partial k_\mu}\rk\f
\frac{\partial}{\partial k_\mu}
\lk
\lk\frac{\partial\omega(k)}{\partial k_\mu}\rk\f
\frac{\partial}{\partial k_\mu} e^{-is\omega(k)}\rk,\ \ \ \mu=1,...,\nu.
\en
Then by the integration by parts formula,
\eqn
&&
\left|
\int_\BR \ov{f(k)}
(\Psi, e^{-is(\gsb -E(\gsb)+\omega(k))}
\tgsb (k)\gr)_{\FFFg} dk \right| \\
&&
\leq
\frac{1}{s^2}
\sum_{j=1}^J
\int_\BR dk \left|
\frac{\partial}{\partial k_\mu}
\lk \frac{\partial\omega(k)}{\partial k_\mu}\rk\f
\frac{\partial}{\partial k_\mu}
\lk \frac{\partial\omega(k)}{\partial k_\mu}\rk\f
\ov{f(k)}\la_j(k) \right| \\
&&\hspace{4cm} \left|
(\Psi, e^{-is(\gsb -E(\gsb))}
(B_j\otimes\i) \gr)_{\FFFg}  \right| \\
&& \leq
\frac{1}{s^2}
\sum_{j=1}^J
\int_\BR dk \left|
\frac{\partial}{\partial k_\mu}
\lk \frac{\partial\omega(k)}{\partial k_\mu}\rk\f
\frac{\partial}{\partial k_\mu}
\lk \frac{\partial\omega(k)}{\partial k_\mu}\rk\f
\ov{f(k)}\la_j(k) \right|\\ 
&&\hspace{6cm} \| \Psi\|_{\FFFg}  \|(B_j\otimes\i) \gr\|_{\FFFg}. 
\enn
Since the integrand of the right-hand side above is integrable for
$f\in C_0^2(\BR\setminus Y)$, we obtain that 
$$
\int_\BR \ov{f(k)}
(\Psi, e^{-is(\gsb -E(\gsb)+\omega(k))}
\tgsb (k)\gr)_{\FFFg} dk \in L^1([0,\infty),ds).$$
Thus (A.4)-(2) follows and the proof is complete. 
\qed
\bc{sinj}
Let $G$ be a nonnegative multiplication operator on $\LR$. 
In addition to (B.1)-(B.5), suppose that 
\eq{53} \sqrt G \la_j/\omega\in\LR,\ \ \ j=1,...,J.
\en 
Then $\gr\in D(\i\otimes\dg(G)^\han)$.
In particular suppose that 
$$ \la_j/\omega\in\LR,\ \ \ j=1,...,J.$$
Then $\gr\in D(\i\otimes  N^\han)$
\ec
\proof 
From \kak{53} it follows that 
\eqn 
&& \int_\BR G(k)\|(\gsb-E(\gsb)+\omega(k))\f\tgsb(k)\gr\|^2 _{\FFFg}
dk\\
&&<\jjj 
\int_\BR G(k)|\la_j(k)/\omega(k)|^2 dk \| (B_j\otimes\i)\gr\|^2_{\FFFg}<\infty.
\enn
Thus the corollary follows from Theorem \ref{g3}. 
\qed

\begin{example}\label{hiro}
Typical examples of $\la_j$, $j=1,...,J$,  
and $\omega$ are 
$$\omega(k)=|k|,\ \ \ \ \la_j=\rho_j/\sqrt\omega, \ \ \ j=1,...,J,$$ 
with some nonnegative functions $\rho_j$ such that 
$\rho_j\in C^2(\BR)$, $\rho_j/\sqrt\omega\in\LR$ and $\rho_j/\omega\in\LR$, $j=1,...,J$. 
Let $\gamma\geq 0$. 
In addition to (B.1) and (B.3), suppose that 
$$\rho_j\omega^\gamma/\omega^{\frac{3}{2}}\in\LR,\ \ \ j=1,...,J.$$
Then $\gr\in D(\i\otimes\dg(\omega^\gamma)^\han)$. 
\end{example}

\subsection{Absence of ground states}
\bt{d4}
Assume (B.1)-(B.5).
Let $G$ be a nonnegative multiplication operator on $\LR$.
Then $\gsb$ has no ground state $\gr$ in $D(\i\otimes \dg(G)^\han)$ such
that
$$
\frac{\sqrt G}{\omega}
 \sum_{j=1}^J(\gr, (B_j\otimes \i)\gr)_\FFFg \la_j  \not \in \LR.$$
\et
\proof
By Theorems \ref{g3} and  \ref{d1} under  identification \kak{d5},
there exists no ground state $\gr$ in
$D(\i\otimes \dg(G)^\han)$ such that
$$\frac{\sqrt G }{\omega}
(\gr, \tgsb(\cdot) \gr)_\FFFg\not\in \LR.$$
Since
$$(\gr, \tgsb(k) \gr)_\FFFg=
\sum_{j=1}^J(\gr, (B_j\otimes \i)\gr)_\FFFg \la_j(k),$$
the theorem follows.
\qed
\bc{d50}
Assume (B.1)-(B.5) and for some $\ell$,  
$$\la_\ell/\omega\not\in\LR.$$
Then $\gsb$ has no ground state $\gr$ in $D(\i\otimes N^\han) $ such that
$(\gr, (B_{\ell}\otimes \i) \gr)>0$
and $(\gr, (B_j\otimes \i) \gr)\la_j \geq 0$ for all $j$ but $j\not=\ell$.
\ec
\proof
Since
$\sum_{j=1}^J(\gr, (B_j\otimes \i)\gr)_\FFFg \la_j\geq
(\gr, B_{\ell}\gr)\la_\ell$,
we have
$$
\frac{1}{\omega}
 \sum_{j=1}^J(\gr, (B_j\otimes \i)\gr)_\FFFg \la_j \not \in \LR.$$
Thus by Theorem \ref{d4} with $G=1$, the corollary follows.
\qed

\begin{example}\label{hiro2} 
Let $\omega(k)=|k|$ and $\la_j=\rho_j/\sqrt\omega$ 
with some nonnegative functions $\rho_j$ such that 
$\rho_j\in C^2(\BR)$, $\rho_j/\sqrt\omega\in\LR$ and $\rho_j/\omega\in\LR$, $j=1,...,J$. 
Suppose (B.1), (B.3) and 
$$\rho_\ell\omega^\gamma/\omega^{3/2}\not\in\LR$$
for some $\gamma\geq 0$. 
Then $\gsb$ has no ground state $\gr$ in 
$D(\i\otimes\dg(\omega^\gamma)^\han)$ such that 
$(\gr, (B_{\ell}\otimes \i) \gr)>0$
and $(\gr, (B_j\otimes \i) \gr)\la_j \geq 0$ for all $j$ but $j\not=\ell$.
\end{example}

\subsection{Higher order regularities} 
In this subsection, we fix a natural number $n$ and consider cores of 
$(\gsb+1)^n$. 
\subsubsection{Cores of  $(\gsb+1)^n$}
We  define $\ad A k B $ by 
$\ad A 0 B := B$ and $\ad A k B := [A, \ad A {k-1} B]$ for  $k\geq 1$. 
If $D$ is an invariant subspace of $A$ and $B$, we have for all 
$\Psi\in D$, 
\eq{ad}
[A^k, B]\Psi 
=\sum_{\ell=1}^k \cc k\ell \ad A \ell B A^{k-\ell}\Psi,\ \ \ 
\ad A k {BC}\Psi 
=\sum_{\ell=0}^k\cc k\ell \ad A \ell  B\ad A {k-\ell} C\Psi. 
\en 
\bi
\item[(B.6)] 
There exists a dense subspace $\di\subset {\cal H}$ such that 
\bi
\item[(1)] $\di\subset D(A)\cap \left[ 
\cap_{j=1}^JD(B_j)\right]$, 
\item[(2)] $A\di\subset \di$ and $B_j\di\subset \di$,  $j=1,...,J$, 
\item[(3)] $A^n\lceil_{\di}$ is essentially self-adjoint, 
\item[(4)] there exist constants $a_k$ and $b_k$ such that 
 for all $\Psi\in\di$ and  $j=1,...,J$, 
$$\|\aa A k {B_j} \Psi\|_\LR 
\leq a_k\|\avv ^{(k+1)/2}
\Psi\|_\LR 
+b_k\|\Psi\|_\LR,\ \ \ 0\leq k\leq n.$$ 
\ei
\item [(B.7)]
$ \omega^k \la_j/\sqrt\omega\in\LR$ and $\omega^k\la_j\in\LR$ for $0\leq k\leq n$ and 
$j=1,...,J$.
\ei
\begin{example}
Let  $A$ and $B_j$, $j=1,...,J$, be bounded. Then 
(B.6) are satisfied with $\di={\cal H}$. 
\end{example}
\begin{example}
\label{e1}
Let ${\cal S}(\BR)$ be the Schwartz space of rapidly dicreasing $C^\infty$ functions on $\BR$ and $V\in{\cal S}(\BR)$ 
be  real-valued.
Let  $A=-\Delta+\beta V$ and $B_j=-i\nabla_j$, $j=1,...,\nu$. 
Then  
(B.6) is  satisfied with $\di={\cal S}(\BR)$ for 
$\beta$ with $|\beta|$ sufficiently small. 
See  Appendix for details.
\end{example}
Let us define  a Hamiltonian $K$ by 
$$K:=\hz+\alpha\gsbi,$$
where 
$$\hz:=\avv \otimes\i+\i\otimes\dg(\omega).$$
The self-adjoint  operator $(K+1)^n$ is defined through the spectral 
theorem, i.e., 
$$(K+1)^n=\int_{[E(\gsb),\infty)} (\lambda-E(A)+1)^n dE(\lambda),$$
where $E(\lambda)$ is the spectral projection associated with $\gsb$. 
Let 
$$\fffi:=
\di\otimes_{\rm alg} \ffff^{C^\infty(\omega)},$$
where $C^\infty(\omega):=\cap_{n=1}^\infty D(\omega^n)$ and 
$\otimes_{\rm alg}$ denotes the algebraic tensor product. 
Since $\gsb$ leaves $\fffi$ invariant, 
it follows that 
$$\fffi\subset \cap _{n=1}^\infty D(\gsb^n),$$
and 
canonical commutation relations for $a(f)$ and $\add(g)$ hold on $\fffi$.
\bt{self}
Suppose (B.1), (B.6) and (B.7).
Then there exists $\alpha_\ast>0$ such that for $\alpha$ with $|\alpha|<\alpha_\ast$, 
$(K+1)^n$ is self-adjoint on $D((\hz+1)^n)$ and 
essentially self-adjoint on any core of $(\hz+1)^n$. In particular 
it is essentially self-adjoint on $\fffi$.
\et
To prove Theorem  \ref{self} we prepare some lemmas. 
\bl{o2}
Suppose (B.1), (B.6) and (B.7).
Then there exist constants $C_\ell$, $\ell=1,...,m$,  such that,  for $\Psi\in\fffi$, 
$$\|[(\hz+1)^m,\gsbi]
\Psi\|_{\FFFg}\leq \sum_{\ell=1}^m\cc m\ell 
C_\ell \|(\hz+1)^{m+1-(\ell/2)}\Psi\|_\FFFg.$$
\el
\proof 
We see that,  for $\Psi\in\fffi$, 
\eqnn 
\label{pop}
[(\hz+1)^m, \gsbi]\Psi
=\sum_{\ell=1}^m \cc m \ell \ad{\hz+1} \ell 
\gsbi (\hz+1)^{m-\ell}\Psi.
\ennn
Using formula \kak{ad} we have 
\eqn 
\ad{\hz+1} \ell \gsbi\Psi 
&=& \jjj\ad \hz \ell {(B_j\otimes\i)(\i\otimes\phi(\la_j)} \Psi  \\
&=& \jjj\sum_{k=0}^\ell\cc \ell k \ad \hz k {B_j\otimes\i} 
\ad \hz {\ell-k} {\i\otimes\phi(\la_j)} \Psi  \\
&=& \jjj\sum_{k=0}^\ell\cc \ell k \ad {\avv } k {B_j} \otimes 
\ad {\dg (\omega)} {\ell-k} {\phi(\la_j)} \Psi  \\
&=& \jjj\sum_{k=0}^\ell\cc \ell k \ad {\avv } k {B_j} \otimes 
\phi((-i)^{\ell-k}\omega^{\ell-k}\la_j) i^{\ell-k}\Psi.
\enn
From   (B.6)-(4), it follows that 
\eqn 
&& \|\ad {\avv } k {B_j} \otimes 
\phi((-i)^{\ell-k}\omega^{\ell-k}\la_j) i^{\ell-k} \Psi\|_{\FFFg}\\
&&\leq a_k\|(\avv ^{(k+1)/2}\otimes \phi((-i)^{\ell-k}\omega^{\ell-k}\la_j))\Psi\|_{\FFFg}+
b_k\|(\i\otimes \phi((-i)^{\ell-k}\omega^{\ell-k}\la_j))\Psi\|_{\FFFg}\\
&&\leq 
\xi_{\ell-k,j}\lkk 
a_k\|(\avv ^{(k+1)/2}\otimes (\dg(\omega) +1)^\han )\Psi\|_{\FFFg}+
b_k\|(\i\otimes (\dg(\omega) +1)^\han)\Psi\|_{\FFFg}\rkk,
\enn
where 
$$\xi_{m,j}:
=(\|\omega^{m}\la_j/\sqrt\omega\|_\LR 
+2\|\omega^{m}\la_j\|_\LR)/\sqrt 2.$$
Note that 
$$
\|(\avv ^{(k+1)/2}\otimes (\dg(\omega) +1)^\han )\Psi\|_{\FFFg}\leq \|(\hz+1)^{(k+2)/2}\Psi\|_{\FFFg},\ \ \ k\geq0,$$
and 
$$
\|(\i\otimes (\dg(\omega) +1)^\han)\Psi\|_{\FFFg}\leq \|(\hz+1)^\han \Psi\|_{\FFFg}.$$
Hence  we have 
\eqn
&&\| \ad {\avv } k {B_j}\otimes \phi((-i)^{\ell-k}\omega^{\ell-k}\la_j))i^{k-\ell}\Psi\|_{\FFFg}\\
&&\leq \xi_{\ell-k,j}\lk a_k\|(\hz+1)^{(k+2)/2}\Psi\|_{\FFFg} + b_k \|(\hz+1)^\han\Psi\|_{\FFFg}\rk.
\enn 
From this it follows that 
\eqnn 
&&
\| \ad{\hz+1} \ell \gsbi\Psi\|_{\FFFg}\nonumber \\
&&
\leq 
\jjj\sum_{k=0}^\ell\cc\ell k 
\xi_{\ell-k,j}  \lk a_k\|(\hz+1)^{(k+2)/2}\Psi\|_{\FFFg} + b_k \|(\hz+1)^\han\Psi\|_{\FFFg}\rk\nonumber \\
&& \label{subb} 
\leq 
\lk 
\sum_{k=0}^\ell\cc\ell k 
\jjj
\xi_{\ell-k,j}   (a_k+b_k) \rk 
\|(\hz+1)^{(\ell+2)/2}\Psi\|_{\FFFg}.
\ennn
Hence from \kak{subb} and \kak{pop} the lemma follows.
\qed
\bl{o1}
Suppose (B.1), (B.6) and (B.7). 
Then there exist constants $c_k$, $k=1,...,n$,  such that  
\eq{k}
\|(\hz+1)^k\gsbi\Psi\|_{\FFFg}\leq c_{k+1} 
\|(\hz+1)^{k+1}\Psi\|_{\FFFg},\ \ \ \Psi\in\fffi,\ \ \ 0\leq k\leq n-1.
\en 
\el
\proof 
We prove the lemma by induction with respect to $k$. 
For $k=0$, \kak{k} holds true. 
Assume that \kak{k} is satisfied for $k=0,1,...,m-1$. 
We see that for $\Psi\in\fffi$, 
$$ 
(\hz+1)^m \gsbi\Psi
= 
\gsbi (\hz+1)^m\Psi+[(\hz+1)^m,\gsbi]\Psi.
$$ 
By Lemma \ref{o2} we have 
\eqn 
&&\|(\hz+1)^m\gsbi\Psi\|_{\FFFg}\\
&&\leq \|\gsbi (\hz+1)^m\Psi\|_{\FFFg}+
\sum_{\ell=1}^m\cc m \ell C_\ell \|(\hz+1)^{m+1-(\ell/2)}\Psi\|_{\FFFg}\\
&&\leq c_0\|(\hz+1)^{m+1}\Psi\|_{\FFFg}
+\sum_{\ell=1}^m\cc m \ell C_\ell \|(\hz+1)^{m+1-(\ell/2)}\Psi\|_{\FFFg}\\
&&\leq 
(c_0+\sum_{\ell=1}^m\cc m \ell C_\ell) \|(\hz+1)^{m+1}\Psi\|_{\FFFg}
\enn
with some constant $c_0$. 
Thus the lemma follows with $c_{m+1}=c_0+\sum_{\ell=1}^m\cc m \ell C_\ell$, $m\geq 1$.
\qed

{\it Proof of Theorem \ref{self}}\\
We have on $\fffi$, 
$$(K+1)^n=(\hz+1)^n+\alpha\hi(n),$$
where 
$$\hi(n):=
 \hi^{(1)}+\alpha\hi^{(2)}+\cdots+\alpha^{n-1}\hi^{(n)},$$
and 
\eqn
&&\hi^{(1)}:=\sum_{i=1}^n \underbrace{
(\hz+1)\cdots \stackrel{i}{\gsbi} \cdots (\hz+1)}_n,\\
&&\hi^{(2)}:=\sum_{i_1<i_2} ^n \underbrace{
(\hz+1)\cdots \stackrel{i_1}{\gsbi} \cdots \stackrel{i_2}{\gsbi} \cdots (\hz+1)}_n,\\
&&\hi^{(3)}:=\sum_{i_1<i_2<i_3} ^n \underbrace{ 
(\hz+1) \cdots \stackrel{i_1}{\gsbi} \cdots 
\stackrel{i_2}{\gsbi} \cdots \stackrel{i_3}{\gsbi} \cdots 
(\hz+1)}_n,\\
&&\hspace{1cm}\vdots\\
&&\hi^{(n)}:=\gsbi ^n.
\enn
We see that 
$$(\hz+1)^n\Psi =
\sum_{k=0}^n \sum_{\ell=0}^k  \cc n k  \cc k \ell\avv ^\ell\otimes \dg (\omega) ^{k-\ell}\Psi,\ \ \ \Psi\in\fffi,$$
and 
$\sum_{k=0}^n \sum_{\ell=0}^k  \cc n k  \cc k \ell\avv ^\ell\otimes \dg (\omega) ^{k-\ell}$ 
is essentially self-adjoint on 
$$C(\avv ^n)\otimes_{\rm alg} C(\dg(\omega)^n),$$
where $C(\avv ^n)$ and $C(\dg(\omega)^n)$ are any 
cores of $ A^n$ and $\dg(\omega)^n$, respectively. 
In particular $\fffi$ is a core of $(\hz+1)^n$. 
From Lemma  \ref{o1} and the definition of $\hi^{(j)}$, $j=1,...,n$,  
we can see that for $\Psi\in\fffi$, 
$$\|\hi^{(j)}\Psi\|_{\FFFg}\leq d_j \|(\hz+1)^n\Psi\|_{\FFFg},\ \ \ j=1,...,n,$$
with some constant $d_j$, which implies that 
$$ \|\hi(n)\Psi\|_{\FFFg}
\leq (d_1+|\alpha| d_2+\cdots +|\alpha|^{n-1} d_n)
\|(\hz+1)^n\Psi\|_{\FFFg}.$$
Since $\fffi$ is a core of $(\hz+1)^n$, 
we can see that 
 $D((\hz+1)^n)\subset D(\ov{\hi(n)})$ and 
$$\|\ov{\hi(n)}\Psi\|_{\FFFg} \leq (d_1+|\alpha| d_2+\cdots +|\alpha|^{n-1} d_n)
\|(\hz+1)^n\Psi\|_{\FFFg},\ \ \ \Psi\in D((\hz+1)^n),$$ 
where $\ov{\hi(n)}$ denotes the closure of $\hi(n)\lceil_{\fffi}$. 
For $\alpha$ such that 
$$|\alpha| (d_1+|\alpha| d_2+\cdots +|\alpha|^{n-1} d_n)<1,$$
by the Kato-Rellich theorem
$$K_n:=(\hz+1)^n+\alpha \ov{\hi(n)}$$
is  self-adjoint on $D((\hz+1)^n)$ and bounded from below. Moreover 
it is essentially self-adjoint on any core of $D((\hz+1)^n)$. 
In particular $K_n\lceil_{\fffi}$ is essentially self-adjoint. 
Let 
$$\alpha_\ast:= \max\lkk |\alpha| \left| 
|\alpha|(
d_1+|\alpha| d_2+\cdots +|\alpha|^{n-1} d_n) \leq 1\right.\rkk.$$
Since 
$$(K+1)^n\lceil_{\fffi}=K_n\lceil_{\fffi}\subset K_n\lceil_{D((\hz+1)^n)}$$ 
and $K_n\lceil_{D((\hz+1)^n)}$ is self-adjoint for $\alpha$ with $|\alpha|<\alpha_\ast$, 
we conclude  that 
$$(K+1)^n=K_n\lceil_{D((\hz+1)^n)},$$ 
i.e.,  $(K+1)^n$ is self-adjoint on $D((\hz+1)^n)$ and 
essentially self-adjoint on any core of $(\hz+1)^n$ for $\alpha$ with $|\alpha|<\alpha_\ast$. 
Thus we get the desired results.
\qed
\subsubsection{Higher order regularities  of ground states} 
\bl{o3}
Suppose (B.1), (B.6) and (B.7).
Then there exist constants $\alpha_{\ast\ast}>0$ and $\xi_k$, $k=1,...,n$,  
such that for $\alpha$ with 
$|\alpha|<\alpha_{\ast \ast}$, 
$$\|(\hz+1)^k \Psi\|_{\FFFg}\leq \xi_k\|(K+1)^k 
\Psi\|_{\FFFg},\ \ \ 1\leq k\leq n,\ \ \ \Psi\in D((K+1)^k).$$
In particular $-1\in \rho(K)$ and 
$(\hz+1)^k (K+1)^{-k}$, $k=1,...,n$,  is a bounded operator with 
$$\|(\hz+1)^k (K+1)^{-k}\|_{\FFFg\rightarrow\FFFg}\leq \xi_k.$$
\el
\proof 
We prove the lemma by induction with respect to $k$.  
For $k=1$, we have 
\eqn 
\|(\hz+1)\Psi\|_{\FFFg}&\leq&
 \|(K+1)\Psi\|_{\FFFg}+|\alpha| 
\|\gsbi\Psi\|_{\FFFg}\\
&\leq& \|(K+1)\Psi\|_{\FFFg}+|\alpha| 
c_0 \|(\hz+1)\Psi\|_{\FFFg}
\enn 
with some constant $c_0$. 
Thus 
\eq{d}
\|(\hz+1)\Psi\|_{\FFFg}\leq \xi_1\|(K+1)\Psi\|_{\FFFg},\ \ \ \Psi\in\fffi,
\en
with $ \xi_1=1/(1-|\alpha| c_0)$  
follows for $\alpha$ with $|\alpha|<1/c_0$. 
Since $\fffi$ is a core of $K+1$, \kak{d} can be extended for $\Psi\in D(K+1)$.
Thus the lemma follows for $k=1$.
Suppose that the lemma holds for  $k=m<n$. 
Note that for $\Psi\in\fffi$,  
\eqn 
&& (\hz+1)^{m+1}\Psi=
(\hz+1)(K+1)\f(\hz+1)^m(K+1)\Psi\\
&& \hspace{3cm} +(\hz+1)(K+1)\f[K+1, (\hz+1)^m]\Psi.
\enn 
We have 
\eq{b1}
\|(\hz+1)(K+1)\f(\hz+1)^m(K+1)\Psi\|_{\FFFg}\leq \xi_1\xi_m\|(K+1)^{m+1}\Psi\|_{\FFFg}
\en 
and by Lemma \ref{o2}, 
\eqn 
&&
\|(\hz+1)(K+1)\f 
[K+1, (\hz+1)^m]\Psi\|_{\FFFg}\\
&& \leq \xi_1|\alpha|\|[\gsbi,(\hz+1)^m]\Psi\|_{\FFFg}\\
&& \leq \xi_1|\alpha| \sum_{\ell=1}^m\cc m\ell C_\ell \|(\hz+1)^{m+1
-(\ell/2)}\Psi\|_{\FFFg}\\
&& \leq 
\xi_1|\alpha| \lk 
\sum_{\ell=2}^m\cc m\ell C_\ell \|(\hz+1)^{m+1-(\ell/2)}\Psi\|_{\FFFg}+m C_1
\|(\hz+1)^{m+(\han)}\Psi\|_{\FFFg}\rk\\
&& \leq 
\xi_1|\alpha| \lk 
\sum_{\ell=2}^m\cc m\ell C_\ell \|(\hz+1)^{m}\Psi\|_{\FFFg}+m C_1
\|(\hz+1)^{m+1}\Psi\|_{\FFFg}\rk.
\enn
Thus 
we have 
\eqnn 
&& \|(\hz+1)^{m+1}\Psi\|_{\FFFg}\nonumber \\
&& \leq 
\frac{\xi_1}{1-\xi_1|\alpha| m C_1} \lk \xi_m 
\|(K+1)^{m+1}
\Psi\|_{\FFFg}+|\alpha| \sum_{\ell=2}^m\cc m\ell C_\ell \|(\hz+1)^m\Psi\|_{\FFFg}\rk\nonumber \\
&&\label{hil}
\leq 
\frac{\xi_1}{1-\xi_1|\alpha| m C_1} (\xi_m + 
|\alpha| \sum_{\ell=2}^m\cc m\ell C_\ell \xi_m) 
\|(K+1)^{m+1}\Psi\|_{\FFFg},\ \ \ \Psi\in\fffi,
\ennn 
for $\alpha$ with $|\alpha|<1/\xi_1m C_1$.  
Since $\fffi$ is a core of $(K+1)^{m+1}$, \kak{hil} can be extended for $\Psi\in D((K+1)^{m+1})$. 
 Thus the lemma follows with $\alpha_{\ast\ast}:=1/(n\xi_1 C_1)$. 
\qed
Let 
$$\ez:=E(A)-1.$$ 
\bc{05}
Suppose (B.1), (B.6) and (B.7).
Then $\ez \in\rho(\gsb)$ and for 
$\alpha$ with $|\alpha|<\min\{\alpha_\ast,\alpha_{\ast\ast}\}$, 
operator 
$(\i\otimes \dg(\omega)^m)(\gsb-\ez)^{-n}$  for $m\leq n$ 
is a bounded operator. 
\ec
\proof 
We have 
\eq{um}
\|(\i\otimes \dg(\omega)^m) \Psi\|_{\FFFg}
\leq 
\|(\hz+1)^m\Psi\|_{\FFFg}
\leq 
\|(\hz+1)^n\Psi\|_{\FFFg}
\en 
for $\Psi\in\fffi$. 
Since $\fffi$ is a core of 
$(\hz+1)^n$, \kak{um} can be extended for 
$\Psi\in D((\hz+1)^n)$. 
Thus $(\i\otimes \dg(\omega)^m)(\hz+1)^{-n}$ 
is a bounded operator with 
$$\|(\i\otimes \dg(\omega)^m)(\hz+1)^{-n}\|_{\FFFg\rightarrow \FFFg}
\leq 1.$$ 
Hence  Lemma \ref{o3} yields that 
\eqn 
&& 
\|
(\i\otimes \dg(\omega)^m)(\gsb-\ez)^{-n}\Psi\|_{\FFFg}\\
&& \leq 
\|(\i\otimes \dg(\omega)^m)(\hz+1)^{-n}(\hz+1)^n (K+1)^{-n}\Psi\|_{\FFFg}\\
&&\leq 
\xi_n\|\Psi\|_{\FFFg}.
\enn 
Thus the corollary  follows. 
\qed
\bl{B}
Suppose (B.1), (B.6) and (B.7). 
Then for $\alpha$ with $|\alpha|<\alpha_\ast$, 
\eq{DD}
\tgsb(k):D((\gsb-\ez)^m)\longrightarrow  
D((\gsb-\ez)^{m-1}),\ \ \ 1\leq m\leq n.
\en 
In particular 
\eq{DDD}
D((\gsb-\ez)^n)\subset D(\tgsb(k_m)^\ast e^{it_m\wgs}\cdots 
\tgsb(k_1) ^\ast e^{it_1\wgs}\nonumber \\
),\ \ \ 1\leq m\leq n,
\en 
where 
$\wgs:=\gsb-E(\gsb)$. 
\el
\proof
Let $\Psi\in\fffi$. 
We have 
\eqn 
&&(\hz+1)^{m-1}(B_j\otimes \i)\Psi \\
&&=(B_j\otimes \i)(\hz+1)^{m-1}\Psi +[(\hz+1)^{m-1},B_j\otimes\i]\Psi\\
&&=(B_j\otimes \i)(\hz+1)^{m-1}\Psi +\sum_{\ell=1}^{m-1}\cc{m-1}\ell \ad \hz \ell {B_j\otimes\i}
(\hz+1)^{m-1-\ell}\Psi.
\enn
By (B.6)-(4), we have 
\eqn 
&& \|\ad \hz \ell {B_j\otimes\i}\Psi\|_\FFFg=\|(\ad  A  \ell {B_j}\otimes\i)\Psi\|_\FFFg\\
&& \leq 
a_\ell\|\lk \avv ^{(\ell+1)/2}\otimes\i \rk \Psi\|_\FFFg+b_\ell\|\Psi\|_\FFFg
\leq c_\ell\|(\hz+1)^{(\ell+1)/2}\Psi\|_\FFFg
\enn 
with some constant $c_\ell$. 
Hence it follows that 
\eqn
&&
\|(\hz+1)^{m-1}(B_j\otimes \i)\Psi\|_\FFFg\\
&&
\leq c_0\|(\hz+1)^{m-(\han)}\Psi\|_\FFFg
+\sum_{\ell=1}^{m-1}\cc {m-1}\ell c_\ell\|(\hz+1)^{m-(\ell+1)/2}\Psi\|_\FFFg\\
&&
\leq 
C\|(\hz+1)^m\Psi\|_\FFFg
\enn
with come constant $C$. 
Thus for $\Psi,\Phi\in\fffi$, it follows that 
\eq{C}
|( (B_j\otimes\i)\Psi, (\hz+1)^{m-1}\Phi) _\FFFg|
\leq C\|(\hz+1)^m\Psi\|_\FFFg \|\Phi\|_\FFFg .
\en 
It is seen that 
$$ 
\|(B_j\otimes\i)
\Psi\|_{\FFFg} \leq C \|(\hz+1)^{1/2}\Psi \|_{\FFFg},\ \ \ \Psi\in D((\hz+1)^\han)).$$
From this it follows that 
\eq{cl}
\|(B_j\otimes\i)
\Psi\|_{\FFFg} \leq C \|(\hz+1)^{m}\Psi \|_{\FFFg},\ \ \ \Psi\in D((\hz+1)^m)).
\en
Since $\fffi$ is a core of $(\hz+1)^m$ and $B_j$  is a closed operator, 
using \kak{cl} we can extend \kak{C} for $\Psi\in D((\hz+1)^m)$ and $\Phi\in D((\hz+1)^{m-1})$.
Set 
$$Q(\Psi, \Phi):=((B_j\otimes\i)\Psi, (\hz+1)^{m-1}\Phi)_\FFFg,\ \ \ 
\Psi\in D((\hz+1)^m), \Phi\in D((\hz+1)^{m-1}).$$
For each fixed $\Psi\in D((\hz+1)^m)$, 
$Q(\Psi, \Phi)$ can be extended for all $\Phi\in\FFFg$ by \kak{C} 
as a linear bounded functional, which is denoted by 
$\ov Q(\Psi, \Phi)$. Thus,   
by the Riesz representation theorem, 
there exists a unique $F_\Psi\in\FFFg$ such that 
\eq{CCC}
\ov Q(\Psi, \Phi)= (F_\Psi, \Phi)_\FFFg,\ \ \ \Psi\in D((\hz+1)^m), \Phi\in\FFFg.
\en 
In particular 
$$
((B_j\otimes\i)\Psi, (\hz+1)^{m-1}\Phi ) _\FFFg=
(F_\Psi, \Phi)_\FFFg,\ \ \ \Psi\in D((\hz+1)^m), \Phi\in D((\hz+1)^{m-1}).
$$ 
This implies that 
$(B_j\otimes\i)\Psi\in D((\hz+1)^{m-1})$ for $\Psi\in D((\hz+1)^m)$, i.e., 
$$B_j\otimes \i :D((\hz+1)^m)\rightarrow D((\hz+1)^{m-1}).$$ 
Since $\tgsb(k)=(\jjj \la_j(k) B_j)\otimes\i$, 
we have $$\tgsb(k):D((\hz+1)^m)\rightarrow D((\hz+1)^{m-1}).$$ 
\kak{DD} follows from the fact that 
$D((\hz+1)^m)=D((K+1)^m)=D((\gsb-\ez)^m)$.
Noting that 
$$e^{it\wgs}:D((\gsb-\ez)^m)\rightarrow D((\gsb-\ez)^m),$$
we can conclude \kak{DDD} and  the proof is complete.  
\qed

\bt{k5}
Suppose (B.1), (B.4)--(B.7) and 
\eq{517} \la_j/\omega\in\LR,\ \ \ j=1,..,J.
\en 
Then for $\alpha$ with $|\alpha|<\min\{\alpha_\ast,\alpha_{\ast\ast}\}$, 
$$\gr\in  D(\i\otimes N^{n/2}).$$
\et
\proof
By Theorem \ref{k4} it is enough to  check (H.1)-(H.5) under  identifcation \kak{d5} and 
$${\cal E}=D((\gsb-\ez)^n),
\ \ \ {\cal C}_n=C_0^2(\RR^\nu\setminus Y).$$
Since 
$\tgsb(k)=\lk\jjj\la_j(k)B_j\rk\otimes \i$,  (H.1) is satisfied.
In Corollary \ref{05} we checked 
that $(\i\otimes\dg(\omega)^m)(\gsb-\ez)^{-n}$, $m\leq n$, 
 is a bounded operator. 
This implies (H.2). 
Let 
$$\beta :=\jjj a_j\|\la_j/\sqrt\omega\|^2_\LR.$$
Then 
\eqn
\|(\avv  ^\han\otimes\i) \Psi\|_\FFFg^2
&\leq &
(\Psi, \gsbz \Psi)_\FFFg+
E(A)\|\Psi\|_\FFFg^2\\
&\leq& 
(\half+|E(A)|) \|\Psi\|_\FFFg^2+\half \|\gsbz\Psi\|_\FFFg^2\\
&\leq& 
(\half+|E(A)|) \|\Psi\|_\FFFg^2+
\frac{1}{2 (1-|\alpha| \beta)^2} \|\gsb\Psi\|_\FFFg^2,
\enn
and 
$$
\|(\avv ^\han \otimes\i)\Psi\|_\FFFg\leq 
\frac{1}{\sqrt 2(1-|\alpha| \beta)} \|\gsb \Psi\|_\FFFg+
\lk {\half+|E(A)|} \rk^\han \|\Psi\|_\FFFg.
$$
Hence we have 
\eqn
&&\|\tgsb(k)\Psi\|_\FFFg\leq 
\jjj\|\la_j(k)(B_j\otimes\i) \Psi\|_\FFFg \\
&&\leq \jjj|\la_j(k)|\lk a_j\|(\avv ^\han\otimes \i)\Psi\|_\FFFg+b_j\|\Psi\|_\FFF\rk\\
&& \leq \jjj|\la_j(k)|\lkk
\frac{a_j/\sqrt 2}{1-|\alpha| \beta}\|\gsb \Psi\|_\FFFg+
\lk b_j+a_j\lk{\half+|E(A)|}\rk^\han\rk 
\|\Psi\|_\FFFg\rkk\\
&&
\leq \jjj|\la_j(k)|
\lk
d_j \|\wgs \Psi\|+
d_j' \|\Psi\|\rk,
\enn 
where 
$$
d_j=
\frac{a_j/\sqrt 2}{1-|\alpha| \beta},\ \ \ 
d'_j=
b_j+a_j\lk {\half+|E(A)|}\rk^\han+ \frac{a_j|E(\gsb )|/\sqrt 2}{1-|\alpha| \beta}.
$$
Thus 
we can obtain that 
$$\|\tgsb(k)(\wgs+\omega(k))\f\Psi\|_\FFFg 
\leq 
\jjj|\la_j(k)|\lk d_j+d_j'\frac{1}{\omega(k)}\rk  \|\Psi\|_\FFFg,$$
from which it follows that 
$$\|[(\wgs+\omega(k))\f\tgsb(k)\ov]\Psi\|_\FFFg \leq 
\jjj|\la_j(k)| \lk d_j+d_j'\frac{1}{\omega(k)}\rk  \|\Psi\|_\FFFg. $$
From (B.2) and \kak{517}, 
it follows that 
$$\sqrt{\omega} \|[(\wgs+\omega(\cdot))\f\tgsb(\cdot)\ov]\Psi\|_\FFFg, 
\|[(\wgs+\omega(\cdot))\f\tgsb(\cdot)\ov]\Psi\|_\FFFg 
\in \LR.$$
Thus 
(H.4) and (H.5) follow.
Finally we shall check  from (H.3)-(1) to (H.3)-(4). 
(H.3)-(1) is   trivial. 
In Lemma \ref{B} we obtained that 
$$D((\gsb-\ez)^n)\subset D(\tgsb(k_m)^\ast e^{it_m\wgs}\cdots 
\tgsb(k_1) ^\ast e^{it_1\wgs}),\ \ \ 1\leq m\leq n,$$
which implies (H.3)-(2). 
Note that 
\eqn
&& \tgsb(k_m)^\ast e^{it_m\wgs}\cdots 
\tgsb(k_1) ^\ast e^{it_1\wgs}\nonumber \\
&& =
\sum_{\ell_1,...,\ell_m}^J 
\ov{\la_{\ell_m}(k_m)}
\cdots 
\ov{\la_{\ell_1}(k_1)}
(B_{\ell_m} \otimes\i) e^{it_m\wgs} \cdots 
(B_{\ell_1} \otimes\i) e^{it_1\wgs},
\ \ \ \enn
Using \kak{ki} and the integration by parts formula,
we obtain that for $\Psi\in D((\gsb-\ez)^n) $, $f_j\in C_0^2(\BR\setminus Y)$, $j=1,...,n$,
 and $T_m=t_1+\cdots+t_m$, 
\eqnn
& &
\left| 
\int_\BR dk_m e^{-i T_m  \omega(k_m)} \ov{f_m(k_m)}\right.\nonumber \\
&&\left.  (\qgsb m   \Psi,  (
\i\otimes a (e^{iT_m \omega}f_{m+1})
\cdots a(e^{iT_m  \omega}f_n)  )\gr)_\FFFg \right|\nonumber 
\\
& &
\leq 
\frac{1}{|T_m|^2}\jjj  \int_\BR dk_{m} 
\left| F_j(k_m)
(\qgsb  {m-1}  \Psi, \right.\nonumber \\
&&\hspace{5cm} \left.  e^{-it_m\wgs} B_j\otimes a (e^{iT_m \omega}f_{m+1})
\cdots a(e^{iT_m  \omega}f_n)  \gr)_\FFFg \right|\nonumber\\
& & 
\leq \frac{1}{|T_m|^2} \jjj \int_\BR dk_{m} 
|F_j(k_m)| 
\left\| 
\qgsb  {m-1}  \Psi\right\|_\FFFg\nonumber\\
&&\label{rhd}
\hspace{5cm} 
\left\| 
 B_j\otimes a (e^{iT_m \omega}f_{m+1})
\cdots a(e^{iT_m  \omega}f_n)  \gr\right\|_\FFFg,
\ennn
where 
$$F_j(k_m)=
\frac{\partial}{\partial {k_{m}}_\mu}
\lk \frac{\partial\omega({k_{m}})}{\partial {k_{m}}_\mu}\rk\f
\frac{\partial}{\partial {k_{m}}_\mu}
\lk \frac{\partial\omega({k_{m}})}{\partial {k_{m}}_\mu}\rk\f
\ov{f_{m}({k_{m}})}\la_j(k_m).$$
Since 
$$
\left\| 
 [B_j\otimes a (e^{iT_m \omega}f_{m+1})
\cdots a(e^{iT_m  \omega}f_n)\ov]  \gr\right\|_\FFFg\leq 
C \left\|(
B_j\otimes \dg(\omega)^{(n-m)/2} )\gr
\right\|_\FFFg
$$ with some $C$ independent of $T_m$. 
Then the integrand of the right-hand side of \kak{rhd} is independent of $T_m$ and 
integrable by  (B.4) and (B.5). 
Thus the right-hand side of \kak{rhd} is in $L^1([0,\infty);dt_m)$. 
Hence (H.3)-(3) follows. 
Let 
$$R^\s_k=(\wgs+\sum_{\ell=k}^n\omega(k_{\s(\ell)}))\f,\ \  \ \s\in{\cal P}_n.$$
Using \kak{ki} and the integration by parts formula again,
we see that \\ 
for $\Psi\in D((\gsb-\ez)^n) $, $f_j\in C_0^2(\BR\setminus Y)$, $j=1,...,n$,
\eqnn 
&&
\left| 
\int_\BR dk_{\s(m)} \cdots \int_\BR dk_{\s(n)}
\prod_{j=m}^n f_{\s(j)}(\ks) \right. \nnn \\
&&
\hspace{2cm} (\qgsbm m \Psi, \nnn \\
&& 
\left. 
\hspace{3cm} \hspace{4cm} R_{m+1}^\s \tgsb (k_{\s(m+1)}) \cdots 
R_{n}^\s \tgsb (k_{\s(n)})\gr)_{\FFFg}\right| \nnn \\
&&
=\left|
\int_\BR dk_{\s(m)} \cdots \int_\BR dk_{\s(n)}
\prod_{j=m}^n f_{\s(j)}(\ks) \right. \nnn \\
&&
\hspace{1cm} 
e^{it_1\omega(k_{\s(1)})} e^{i(t_1+t_2)\omega(k_{\s(2)})} \cdots 
e^{i(t_1+\cdots +t_{m-1})\omega(k_{\s(m-1)})}
e^{i(t_1+\cdots +t_m)(\omega(k_{\s(m)}+\cdots +\omega(k_{\s(n)})} \nnn \\
&&
\hspace{3cm} 
(\tgsb(k_{\s(m)})^\ast e^{it_{m}\wgs} \cdots \tgsb(k_{\s(1)}^\ast e^{it_1\wgs}
 \Psi, \nnn \\
&& 
\left. 
\hspace{3cm} \hspace{4cm} R_{m+1}^\s \tgsb (k_{\s(m+1)}) \cdots 
R_{n}^\s \tgsb (k_{\s(n)})\gr)_{\FFFg}\right| \nnn \\
&&
=\frac{1}{t_1+\cdots+t_m}
\left|
\jjj \int_\BR dk_{\s(m)} \cdots \int_\BR dk_{\s(n)}
\prod_{j=m+1}^n f_{\s(j)}(\ks) F_j(k_{\s(m)})  \right.\nnn \\
&&
\hspace{1cm}  e^{it_1\omega(k_{\s(1)})} e^{i(t_1+t_2)\omega(k_{\s(2)})} \cdots 
e^{i(t_1+\cdots +t_{m-1})\omega(k_{\s(m-1)})} 
e^{i(t_1+\cdots +t_m)(\omega(k_{\s(m+1)}+\cdots +\omega(k_{\s(n)})} 
\nnn \\
&&
\hspace{3cm} 
(\tgsb(k_{s(m-1)})^\ast e^{it_{m-1}\wgs} \cdots \tgsb(k_{\s(1)})^\ast e^{it_1\wgs}
 \Psi, \nnn \\
&& 
\left. 
\hspace{4cm}  e^{-it_m\wgs}(B_j\otimes\i) R_{m+1}^\s \tgsb (k_{\s(m+1)}) \cdots 
R_{n}^\s \tgsb (k_{\s(n)})\gr)_{\FFFg}\right|
\nnn \\
&&\leq 
\frac{1}{t_1+\cdots+t_m}
\jjj \int_\BR dk_{\s(m)} \cdots \int_\BR dk_{\s(n)}
\left| 
\prod_{j=m+1}^n f_{\s(j)}(\ks)\right|  |F_j(k_{\s(m)})| 
\nnn \\
&&
\hspace{3cm}
\| \tgsb(k_{s(m-1)})^\ast e^{it_{m-1}\wgs} \cdots \tgsb(k_{\s(1)})^\ast e^{it_1\wgs}
 \Psi\|_\FFFg  \nnn \\
&& 
\label{tiger}
\hspace{3cm} \| (B_j\otimes\i) R_{m+1}^\s \tgsb (k_{\s(m+1)}) \cdots 
R_{n}^\s \tgsb (k_{\s(n)})\gr)_{\FFFg}\|_\FFFg, 
\ennn 
where 
$$F_j(k_{\s(m)})=
\frac{\partial}{\partial {k_{\s(m)}}_\mu}
\lk \frac{\partial\omega({k_{\s(m)}})}{\partial {k_{\s(m)}}_\mu}\rk\f
\frac{\partial}{\partial {k_{\s(m)}}_\mu}
\lk \frac{\partial\omega({k_{\s(m)}})}{\partial {k_{\s(m)}}_\mu}\rk\f
\ov{f_{\s(m)}({k_{\s(m)}})}\la_j(k_{\s(m)}).$$
Then the right-hand side of \kak{tiger} is in $L^1([0,\infty);dt_m)$. Hence (H.3)-(4) follows.
Thus the proof is complete. 
\qed

\section{Concluding Remarks}
In this section we give some remarks on ground states of 
the Nelson model and the Pauli-Fierz model.  
Through this section we assume that $\omega$ is the multiplication operator on $\LR$ 
by 
\eq{lk}
\omega(k):=|k|.
\en 
\subsection{The Nelson models}
The so-called Nelson model was introduced by Nelson \cite{ne3}, which describes 
an iteraction between nonrelativistic 
particles and a scalar quantum field.  
Here we consider only the case where one nonrelativistic particle interacts with a scalar quantum field in $\BR$. Then the  Hilbert space for the Nelson model is defined by
$$\FFFn:=\LR\otimes \fff(\LR)\cong \int^\oplus_\BR \fff(\LR) dx,$$
where 
$\int_{\BR}^\oplus\cdots dx$ denotes a constant fiber direct integral \cite{rs4}. 
The Nelson Hmailtonian, $\n$,  is defined by
$$\n:=\nz+g \Ni,$$
where  $g\in\RR$ is a coupling constant,  
$$
\nz:= \lk-\half\Delta+V\rk \otimes \i+\i\otimes\dg(\omega)
$$
with $V:\BR\rightarrow \RR$ an external potential, 
and 
$$
\Ni:=\int^\oplus_\BR \phi(x) dx
$$ with 
$$ 
\phi(x):= 
\frac{1}{\sqrt 2}\lkk a(\fox) +\add (\fox) \rkk.
$$ 
Here for each $x\in\BR$ we define $\fox\in\LR$ by 
$$\fox(k) :=\la(k)  e^{-ikx}/\sqrt{\omega(k)}.$$ 
\bp{nelson self}
Assume that 
$ \la/\sqrt\omega, \la/\omega\in\LR$ and that $V$ is relatively bounded with respect to 
$\displaystyle -\half \Delta$ with a relative bound strictly less than one. 
Then for all $g\in\RR$, $\n$ is self-adjoint on $D(\nz)$ and bounded from below. 
Moreover it is essentially self-adjoint on any core of $\nz$. 
\ep \proof See \cite{ne3}.
\qed
We see that
$$[\i\otimes a(f),\Ni]_W^{D(\n)}(\Psi,\Phi)=\int_\BR \bar{f}(k)(\Psi,
\tn(k)\Phi)_{\FFFn} dk,$$
where
$$\tn(k):=\frac{\la(k)}{\sqrt{2\omega(k)} }\int_\BR^\oplus e^{-ikx} dx.$$
Under the following identifications 
$$ 
\wh =\n,\ \ \ 
 \hi=\Ni,\ \ \ 
\fs(k)=\omega(k),\ \ \
T(k)=\tn(k), \ \ \ 
\cd=C_0^2(\BR\setminus Y),
$$ 
We can check that $\n$ satisfies assumptions (A.1)-(A.5).
We introduce an assumption. 
\bi
\item[(${\rm IR}$)]
On a neighboorhood of $\{0\}$, $\la$ is continuous and 
$\la(k)\sim |k|^p$, where $2p\leq 3-\nu$.
\ei
Suppose  (${\rm IR}$). Then 
\eq{nnn}
\la/\omega^{3/2}\not\in \LR,
\en 
and if a ground state $\gr$ of $\n$ exists, then  
\eq{nn}
\frac{1}{\omega} 
{(\gr, \tn(\cdot )\gr)_\FFFn}=(\gr, e^{-i\cdot  x} \gr)_\FFFn \frac{\la}{\sqrt 2\omega^{3/2}}
\not\in\LRR.
\en 
Thus it follows from Theorem \ref{d1}  that, if \kak{nnn} holds, then  
$\n$ has no ground state $\gr$ in $D(\i\otimes N^\han)$. 
Actually the absence of ground states of $\n$ under condition 
\kak{nnn} has been established. 
See \cite{h0, lomisp}. 

\subsection{Infrared regular representation of the Nelson models}
The Nelson model in a non-Fock representation is 
introduced and investigated in \cite{a2}. 
The Nelson Hamiltonian  in a non-Fock representation 
is given as a self-adjoint operator on $\FFFn$ by 
$$\nr:=\nz+g\nri,$$ 
where 
$\nri:= \nrii-g W\otimes\i+gc,$ 
and 
$$ 
W(x):=\int_\BR\frac{\la(k)^2}{\omega(k)^2} e^{-ikx} dk,\ \ \ 
c:=\half\|\la/\omega\|_\LR^2,\ \ \ 
\nrii:=\int_\BR^\oplus \phi_{\rm reg}(x) dx.
$$
Here 
$$\phi_{\rm reg}(x):=
\frac{1}{\sqrt2}
\lkk 
\add(\fox-\foo)+ 
a(\fox-\foo)\rkk.$$ 
It is known \cite{a2} that in the case of 
$\la/\omega^{3/2}\in\LR$,  
there exists a unitary operator ${\cal U}$ on $\FFFn$ such that 
\eq{uni}
{\cal U} \n {\cal U}\f= \nr.
\en 
However,  in the case of $\la/\omega^{3/2}\not\in\LR$,  $\n$ and $\nr$ are {\it not} unitarily equivalent. 
We see that 
$$[\i\otimes a(f), \nri]_W^{D(\nr)}(\Psi, \Phi)=\int_\BR\ov{f}(k)(\Psi, 
\tnr(k)\Phi)_\FFFn dk,$$ 
where 
$$\tnr(k):=\frac{\la(k)}{\sqrt{2\omega(k)} }\int_\BR^\oplus (e^{-ikx}-1) dx.$$
Suppose 
that a ground state $\gr$ of $\nr$ exists 
and 
\eq{ex}
\|(|x|\otimes\i)\gr\|_\FFFn<\infty.
\en 
Actually for some $V$, e.g., $V(x)=-1/|x|, |x|^2$, \kak{ex} has been established. 
See, e.g., \cite{bfs1, gll, hi27}. 
Then 
\eqn \frac{1}{\omega(k)} | (\gr, \tnr(k) \gr)_\FFFn| 
&\leq& \frac{\la(k)|k|}{\sqrt 2 \omega(k)^{3/2}} 
\|\gr\|_\FFFn\|(|x|\otimes\i)\gr\|_\FFFn\\
&=& 
\frac{\la(k)}{\sqrt{2\omega(k)}} \|\gr\|_\FFFn\|(|x|\otimes\i)\gr\|_\FFFn.
\enn 
Hence   
\eq{kol}
\frac{1}{\omega} (\gr, \tnr(\cdot) \gr)_\FFFn \in \LR.
\en 
\br{ner}
We do not assume $\la/\omega^{3/2}\in\LR$ in \kak{kol}.
\er
In \cite{a2},  
the existence of a ground state $\gr$ of $\nr$ such that 
$\gr\in D(\i\otimes N^\han)$ is established without assuming $\la/\omega^{3/2}\in\LR$.

\subsection{The Pauli-Fierz models}
The Pauli-Fierz model \cite{pf} 
describes an interaction between nonrelativistic particles and a 
quantum 
radiation field.  
The Hilbert space of the Pauli-Fierz model is given by 
$$\FFFp:=\LR\otimes \fff(\oplus^{\nu-1} \LR)\cong\int_\BR^\oplus 
\fff(\oplus^{\nu-1} \LR)dx.$$
The creation operator and the annihilation operator are denoted by 
$\add (f_1\oplus\cdots\oplus f_{\nu-1})$  
and 
$a(f_1\oplus\cdots\oplus f_{\nu-1})$, respectively.
The Pauli-Fierz Hamiltonian is defined by 
$$ 
\PF:= 
\pfz+e\pfi,
$$ 
where 
\eqn 
&& \pfz:=\lk-\half\Delta+V\rk\otimes\i+\i\otimes\dg(\oplus^{\nu-1}\omega),\\
&&\pfi:=-(p\otimes\i) \cdot A+\frac{e}{2}A\cdot A.
\enn 
Here $p=(-i\partial/\partial x_1,...,-i\partial/\partial x_n)$ denotes 
the set of generalized partial differential operator, $e\in\RR$ 
a coupling constant, 
$V$ an external potential 
and 
$A=(A_1,...,A_\nu)$ denotes a quantum  radition field defined by 
$$A_\mu:=\int_\BR^\oplus A_\mu(x) dx,\ \ \ \mu=1,...,\nu,$$
where 
$$A_\mu(x):=\frac{1}{\sqrt 2} \lkk 
\add(\oplus_j^{\nu-1}e_\mu^j \fox)
+a(\oplus_j^{\nu-1}e_\mu^j \fox)\rkk $$
and $e^j(k)=(e^j_1(k),...,e_\nu^j(k))$ , $j=1,..,\nu-1$, denote 
 $\nu$-dimensional polarization vectors  
such that 
$$e^j(k)\cdot e^{j'}(k)=\delta_{jj'}1,\ \ \ k\cdot e^j(k)=0,\ \ \ j,j'=1,..,\nu-1.$$
We can take $e^1,e^2,...,e^{\nu-1}$ such that 
$e_\mu^j$ ($\mu=1,...,\nu, j=1,...,\nu-1$) 
is continuous on $\BR\setminus Z$ for some $Z\subset \BR$ 
with Lebesgue measure $|Z|=0$. 
\bp{se2}
Suppose that $\sqrt\omega\la,\la,\la/\sqrt\omega, \la/\omega\in\LR$ and that 
external potential $V$ is relatively bounded with respect to 
$\displaystyle -\half \Delta$ with a relative bound strictly less than one.
Then $\PF$ is self-adjoint on 
$D(\pfz)$ for all $e\in\RR$. 
\ep
\proof See \cite{h11,h16}. 
\qed 
We see that 
$$[\i\otimes a(f_1\oplus\cdots\oplus f_{\nu-1}), \pfi]_W^{D(\PF)}=
\sum_{j=1}^{\nu-1} 
\int_\BR\ov{f_j}(k)(\Psi, \tpf_j(k)\Phi)_{\FFFp} dk,$$
where 
$$\tpf_j(k):=
-\frac{\la(k)}{\sqrt{2\omega(k)}} u_k 
 e^j(k) \cdot (p\otimes\i-eA) $$
and $u_k:\FFFp\rightarrow \FFFp$ is the unitary operator defined by $u_k
=\int^\oplus_\BR e^{-ikx} dx$.
Under the following  identifications 
\eqn 
&& 
\wh =\PF,\ \  \hi=\pfi,\ \ \ 
\fs(k)=\oplus^{\nu-1}\omega(k),\ \ \ \\
&& T(k)=\oplus_{j=1}^{\nu-1}\tpf_j(k), \ \ \ \cd=C_0^2(\BR\setminus (Z\cup Y)),
\enn 
we can check that $\PF$ satisfies assumptions (A.1)-(A.5).
Suppose that there exists a ground state $\gr$ of $\PF$ such that 
$\|(|x|\otimes\i)\gr\|_{\FFFp}<\infty$. 
Then we obtain that 
$$ 
\tpf_j(k)\gr=
i\frac{\la(k)}{\sqrt{2\omega(k)}} 
\lkk (\PF-E(\PF))+(p\otimes \i-eA)\cdot k+\half|k|^2\rkk 
u_k e^j(k) \cdot (x\otimes\i)\gr.
$$ 
Then 
\eqnn
&& 
\sum_{j=1}^{\nu-1} (\gr, \tpf_j(k)\gr)_{\FFFp}\nnn \\
&&\label{66}
= 
\sum_{j=1}^{\nu-1} i\frac{\la(k)}{\sqrt{2\omega(k)}}  
(\gr, \lkk (p\otimes \i-eA)\cdot k+\half|k|^2\rkk 
u_k e^j(k) \cdot (x\otimes\i)\gr)_{\FFFp}.
\ennn  
By \kak{66} we can obtain that 
\eq{67}
|\sum_{j=1}^{\nu-1} (\gr, \tpf_j(k)\gr)_{\FFFp}|\leq (c_1|k|+c_2|k|^2)
\|\gr\|_{\FFFp} 
\|(|x|\otimes\i)\gr\|_{\FFFp} 
\en 
with some constants $c_1$ and $c_2$. See \cite{hi27} for details. 
From this we can coclude that 
\eq{pl}
\frac{1}{\omega} 
\sum_{j=1}^{\nu-1} 
(\gr, \tpf_j(k)\gr)_{\FFFp}\in\LR.
\en 
\br{nerr}
We do not assume $\la/\omega^{3/2}\in\LR$ in \kak{pl}.
\er
In \cite{bfs3} the existence of a ground state $\gr$ of $\PF$ 
such that $\gr\in D(\i\otimes N^\han)$ is actually established 
without assuming $\la/\omega^{3/2}\in\LR$. 
\begin{remark}
\kak{pl}  holds for the dipole approximation of $\PF$, too. 
We omit  the details. See \cite{h00}.
\end{remark}

\section{Appendix}
\subsection{Proof of Lemma \ref{1}}
{\it Proof of $(1)\Longrightarrow (2).$}\\
Let $\Psi=\{\Psi^{(n)}\}_{n=0}^\infty \in\ffff(\kkk)$ be such that
$\Psi^{(n)}=\add(f_1\nn)\cdots \add(f_n\nn)\Omega$. 
Then,  using  canonical commutation relations \kak{c1}--\kak{c3},
we can show that
\eqn
&&
\mbox{\rm s-}\!\lim_{M\rightarrow \infty} \sum_{m=1}^M
\add(\T^\ast e_m)a(\T^\ast e_m) \Psi^{(n)}\\
&&=
\mbox{\rm s-}\!\lim_{M\rightarrow \infty}
\sum_{j=1}^n\add(f_1\nn)
\cdots \add( \T^\ast \sum_{m=1}^M (e_m, \T f_j\nn)  e_m)\cdots \add(f_n\nn)
\Omega\\
&&
=\sum_{j=1}^n\add(f_1\nn)
\cdots \add(\T^\ast \T f_j\nn)
\cdots \add(f_n\nn) \Omega\\
&&=\dg(\T^\ast \T)\Psi^{(n)}.
\enn
Hence we have
\eqn
&&\sum_{m=1}^\infty \|a(\T^\ast e_m)\Psi\|_{\fff(\kkk)}^2
=
\sum_{n=0}^{\rm finite} (\Psi^{(n)},
\sum_{m=1}^\infty\add(\T^\ast e_m)a(\T^\ast e_m)\Psi^{(n)})_\k{n}\\
&&
=(\Psi, \dg (\T^\ast \T)\Psi)_{\fff(\kkk)}
=\|\dg (T^\ast \T)^\han\Psi\|_{\fff(\kkk)}^2.
\enn
Since the finite linear hull of such $\Psi$'s, say ${\cal D}$,
is a core of $\dg (\T^\ast \T)^\han$,
we can choose $\Psi_\epsilon\in {\cal D}$ for $\Psi\in D(\dg (\T^\ast
\T)^\han)$
 such that
$\Psi_\epsilon\rightarrow \Psi$ and $\dg(\T^\ast \T)^\han
\Psi_\epsilon\rightarrow \dg(\T^\ast \T)^\han \Psi$ as $\epsilon\rightarrow
0$
strongly.
From the  facts that  $a(f)$ is a closed operator and that by \kak{2},
$$\|a(\T^\ast e_m)\Psi\|_{\fff(\kkk)}
\leq \|\dg (\T^\ast \T)^\han \Psi\|_{\fff(\kkk)},$$
it follows that
$$\lim_{\epsilon\rightarrow 0}
\|a(\T^\ast e_m)\Psi_\epsilon\|_{\fff(\kkk)}
=\|a(\T^\ast e_m)\Psi\|_{\fff(\kkk)}.$$
Then we obtain that
\eqn
&& \sum_{m=1}^M \|a(\T^\ast e_m)\Psi_\epsilon\|_{\fff(\kkk)}^2
\leq \sum_{m=1}^\infty \|a(\T^\ast e_m)\Psi_\epsilon\|_{\fff(\kkk)}^2=
\|\dg(\T^\ast \T)^\han \Psi_\epsilon\|_{\fff(\kkk)}^2,
\enn
and as $\epsilon\rightarrow 0$ on the both sides above,
$$
\sum_{m=1}^M \|a(\T^\ast e_m)\Psi\|_{\fff(\kkk)}^2
\leq
\|\dg(\T^\ast \T)^\han \Psi\|_{\fff(\kkk)}^2.
$$
Hence, taking $M\rightarrow \infty$ on the both sides above,
we can conclude that
$$\sum_{m=1}^\infty  \|a(\T^\ast e_m)\Psi\|_{\fff(\kkk)}^2< \infty.$$
\noindent
{\it Proof of  $(1)\Longleftarrow (2).$}\\
Let $\Psi=\{\Psi^{(n)}\}_{n=0}^\infty$.
We have
\eqnn
&&\hspace{-1cm}
\sum_{m=1}^\infty \|a(\T^\ast e_m)\Psi\|_{\fff(\kkk)}^2 = 
\sum_{m=1}^\infty \sum_{n=1}^\infty \|a(\T^\ast e_m) \Psi^{(n)}\|_\k{n-1}^2
\nonumber \\
&& \hspace{-1cm}=  \lim_{M\rightarrow\infty} \sum_{n=1}^\infty \sum_{m=1}^M \|a(\T^\ast
e_m)
\Psi^{(n)}\|_\k{n-1}^2
= \label{c4}
\sum_{n=1}^\infty \lim_{M\rightarrow\infty}\sum_{m=1}^M  \|a(\T^\ast e_m)
\Psi^{(n)}\|_\k{n-1}^2.
\ennn
Here on the third equality we used the monotone convergence theorem.
The restriction  $A_M:=\sum_{m=1}^M\add(\T^\ast e_m)a(\T^\ast
e_m)\lceil_{\k{n}}$
is a bounded operator, and
$$\|A_M\|_{\k{n}\rightarrow \k{n}}
\leq
\| \dg_n (\T ^\ast \T )^\han\|_{\k{n}\rightarrow \k{n}}.$$
Then
$$\mbox{\rm s-}\!
\lim_{M\rightarrow \infty}A_M=\dg_n (\T ^\ast \T )$$
on $\otimes_s^n\kkk$.
Hence we have by \kak{c4},
\eqn
&& \infty>\sum_{m=1}^\infty \|a(\T ^\ast e_m)\Psi\|_{\fff(\kkk)}^2 = 
\sum_{n=1}^\infty \lim_{M\rightarrow \infty} (\Psi^{(n)},
A_M\Psi^{(n)})_\k{n}\\
&& = \sum_{n=1}^\infty (\Psi^{(n)}, \dg_n (\T ^\ast \T )\Psi^{(n)})_\k{n}
=  \sum_{n=0}^\infty \|\dg_n (\T ^\ast \T )^\han\Psi^{(n)}\|_{\k{n}}^2\\
&& =  \|\dg (\T ^\ast \T )^\han\Psi\|_{\fff(\kkk)}^2.
\enn
Thus the lemma is proven. \qed
\subsection{Proof of Lemma \ref{k1}}
It is seen that for $\Psi=\add(f_1)\cdots \add(f_m)\Omega$,
$$\lim_{m_1,...,m_n\rightarrow\infty}
\sum_{i_1,...,i_n=1}^{m_1,...,m_n}  \|a(e_{i_1})\cdots a
(e_{i_n})\Psi\|_{\fff(\kkk)}^2=
m(m-1)\cdots (m-n+1)\|\Psi\|_{\fff(\kkk)}^2.$$
Hence by a limiting argument we have
for $\Psi\in\ffff(\kkk)$,
$$\lim_{m_1,...,m_n\rightarrow\infty}
\sum_{i_1,...,i_n=1}^{m_1,...,m_n}  \|a(e_{i_1})\cdots a
(e_{i_n})\Psi\|_{\fff(\kkk)}^2=
\|\prod_{j=1}^n (N-j+1) \Psi\|_{\fff(\kkk)}^2.$$
{\it Proof of} $(1) \Longrightarrow (2)$ \\
Let $\Psi_M\in\ffff(\kkk)$ be a truncated vector for $\Psi$
defined by   $\Psi_M^{(m)}:=\lkk\begin{array}{ll} \Psi^{(m)},& m\leq
M,\\0,&m>M.
\end{array}\right.$
 Then
\eqn
\sum_{i_1,...,i_n=1}^{\infty}   \|a(e_{i_1})\cdots a
(e_{i_n})\Psi_M\|_{\fff(\kkk)}^2
 &=&
\|\prod_{j=1}^n (N-j+1) \Psi_M\|_{\fff(\kkk)}^2\\
&=&\sum_{m=0}^M
\|\prod_{j=1}^n (N-j+1) \Psi^{(m)}\|^2_{\k{m}}.
\enn
Take $M\rightarrow \infty$ on the both sides above. Then we have by the
monotone convergence theorem,
\eqn \infty>\sum_{i_1,...,i_n=1}^{\infty}
\|a(e_{i_1})\cdots a (e_{i_n})\Psi\|_{\fff(\kkk)}^2
&=& \sum_{m=0}^\infty
\|\prod_{j=1}^n (N-j+1) \Psi^{(m)}\|^2_{\k{m}}\\
&=&
\|\prod_{j=1}^n (N-j+1) \Psi\|^2_{\fff(\kkk)}.
\enn
Thus (2) follows.\\
{\it Proof of } $(2) \Longrightarrow (1)$
$$
\sum_{i_1,...,i_n=1}^{m_1,...,m_n}  \|a(e_{i_1})\cdots a
(e_{i_n})\Psi\|_{\fff(\kkk)}^2
\leq \|\prod_{j=1}^n (N-j+1) \Psi\|^2_{\fff(\kkk)}.
$$Take $m_1,...,m_n\rightarrow\infty$ on the both sides above. Thus (1)
follows.
\qed
\subsection{Example \ref{e1}}
In this section we shall prove that 
$A=-\Delta+\beta V$, $V\in\S$,  and $B_j=-i\nabla_j(=p_j)$, $j=1,...,\nu$, 
satisfy (B.6) with $\di={\cal S}(\BR)$ for $\beta$ 
with $|\beta|$ sufficiently small.  
\begin{proposition}
\label{e2}
Suppose that $|\beta|$ is sufficiently small. Then 
(1) $A^n$ is self-adjoint on 
$$D(A^n)=D(\D^n)$$ 
and essentially self-adjoint on any core of $\D^n$. 
In particular $A^n$ is essentially self-adjoint on $\S$, 
(2) there exist constants $a_k$ and $b_k$ such that 
 for all $\Psi\in\S$ and  $j=1,...,\nu$, 
$$\|\aa A k {p_j} \Psi\|_\LR 
\leq a_k\|\avv ^{(k+1)/2}
\Psi\|_\LR 
+b_k\|\Psi\|_\LR,\ \ \ k\geq 0.$$ 
\end{proposition}
Before going to a proof of Proposition \ref{e2} we prepare some lemmas. 
In this section we write $\|\cdot \|$ for $\|\cdot\|_\LR$ for simplicity. 
Note that 
$$\|\p 1\cdots\p m\Phi\|\leq \|\D^{m/2}\Phi\|,\ \ \ \Phi\in\S,\ \ \ 1\leq j_1,...,j_m\leq \nu,$$
and 
for $k\leq l$, 
\eq{f1}
\|\D^{k/2}\Phi\|\leq C_{k,\ell}(\|\D^{l/2}\Phi\|+\|\Phi\|),\ \ \ \Phi\in\S, 
\en 
with some constant $C_{k,\ell}$.
The operators $A$ and $p_j$ leave $\S$ invariant 
and 
we see that 
$$A^n\lceil_{\S}=(\D^n+\beta\ihh)\lceil_{\S},$$
where 
\eqn
\ihh&:=& \sum_j\underbrace{
\D\cdots \stackrel{j}{V}\cdots\D}_n
+\beta \sum_{j_1<j_2}\underbrace{ 
\D\cdots \stackrel{j_1}{V}\cdots \stackrel{j_2}{V}\cdots \D}_n\\
&& +\beta^2 \sum_{j_1<j_2<j_3}
\underbrace{ \D\cdots \stackrel{j_1}{V}\cdots \stackrel{j_2}{V}
\cdots \stackrel{j_3}{V}\cdots \D}_n+\cdots
+ \beta^{n-1} V^n.
\enn
\bl{f2}
There exists a constant $C_m$ such that 
$$\|[\p1 \p 2\cdots \p m,V]\Phi\|\leq C_m
(\|\D^{(m-1)/2}\Phi\|+\|\Phi\|),\ \ \ \Phi\in\S, \ \ \ 1\leq j_1,...,j_m\leq \nu.$$
\el
\proof 
Since on $\S$, 
$$[\p1 \p 2\cdots \p m,V]
=\sum_{\ell=1}^m
\sum_{\{i_1,...,i_\ell\}\subset\{j_1,...,j_m\}}
V^{i_1,...,i_\ell}\p1\cdots \sl{\pp 1}\cdots \sl{\pp \ell}\cdots \p m,$$
where 
$$V^{i_1,...,i_\ell}=(-i)^\ell \frac{\partial^\ell V}{\partial 
x_{i_1}\cdots \partial x_{i_\ell}},$$ 
we have 
\eq{71}
\|[\p1 \p 2\cdots \p m,V]\Phi\|
\leq 
\sum_{\ell=1}^m\sum_{\{i_1,...,i_\ell\}\subset\{j_1,...,j_m\}}
\|V^{i_1,...,i_\ell}\|_\infty \|\D^{(m-\ell)/2}\Phi\|.
\en
Here $\|f\|_\infty:= {\rm ess. sup}_{k\in\BR} |f(k)|$. 
From \kak{f1} it follows that 
\eq{72}
\|\D^{(m-\ell)/2}\Phi\|\leq C_{(m-\ell)/2,(m-1)/2}
(\|\D^{(m-1)/2}\Phi\|+\|\Phi\|).
\en 
Then the lemma follows from \kak{71} and \kak{72}. 
\qed 
\bl{f3}
There exists a constant $C_\ell$ such that 
\eq{maru2}
\|\D^\ell V\Phi\|\leq C_\ell (\|\D^\ell\Phi\|+\|\Phi\|),\ \ \ \Phi\in\S.
\en 
\el
\proof 
Note that for $\Phi\in\S$, 
$$\|\D^\ell V\Phi||\leq \|V\D^\ell\Phi\|+\|[\D^\ell,V]\Phi\|.$$
Since on $\S$, 
$$[\D^\ell,V]= \sum_{j_1,...,j_\ell=1}^\nu[\p 1^2\cdots \p \ell^2,V],$$
by Lemma \ref{f2} 
\eq{we1}
\|[\D^\ell,V]\Phi\|\leq C 
(\|\D^{(2l-1)/2}\Phi\|+\|\Phi\|)  \leq 
C'(\|\D^\ell \Phi\|+\|\Phi\|)  
\en 
with some constants $C$ and $C'$. Moreover 
\eq{we2}
\| V\D^\ell\Phi\|\leq\|V\|_\infty\|\D^\ell\Phi\|.
\en 
From \kak{maru2}, \kak{we1} and \kak{we2}, 
it follows that 
$$\|\D^\ell V\Phi\|\leq (C'+\|V\|_\infty)(\|\D^\ell\Phi\|+\|\Phi\|).$$
Then the proof is complete. 
\qed
\bl{f4}
There exists a constant $C_{i_1,...,i_m}$ such that for $1\leq i_1<\cdots<i_m\leq n$, 
$$\|\underbrace{
\D\cdots \stackrel{i_1}{V}\cdots \stackrel{i_2}{V}\stackrel{\cdots}{\cdots} \stackrel{i_m}{V}\cdots \D}_n\Phi\|
\leq C_{i_1,...,i_m}(\|\D^{n-m}\Phi\|+\|\Phi\|),\ \ \ \Phi\in\S.$$
\el
\proof 
It inductively follows from Lemma \ref{f3}. 
\qed
{\it Proof of Proposition \ref{e2} (1)} \\
From the definition of $\ihh$, \kak{f1} and Lemma \ref{f4},  it follows that 
\eqnn
\|\ihh\Phi\|
&\leq& \sum_{\ell=1}^n |\beta|^{\ell-1} C_{\rm I, \ell}
 (\|\D^{n-\ell}\Phi\|+\|\Phi\|)\nonumber \\
\label{f5} 
&\leq& C_{\rm I}( \|\D^n\Phi\|+\|\Phi\|),\ \ \ \Phi\in\S,
\ennn
with some constants $C_{\rm I}$ and $C_{\rm I,\ell}$, $\ell=1,...,n$. 
Since $\S$ is a core of $\D^n$, we can extend \kak{f5} for 
$\Phi\in D(\D^n)$ with 
\eq{f6}
\|\ov{\ihh}\Phi\|\leq C_{\rm I}( \|\D^n\Phi\|+\|\Phi\|),
\en 
where $\ov\ihh$ denotes the closure of $\ihh\lceil_{\S}$. 
Then for $\beta$ with $|\beta|<1/C_{\rm I}$, the  Kato-Rellich theorem yields that 
$\D^n+\beta\ov\ihh$ is self-adjoint on $D(\D^n)$ and bounded from below. Moreover it is 
essentially self-adjoint on any core of $\D^n$. 
In particular $\D^n+\beta \ov\ihh$ is essentially self-adjoint on $\S$. 
Since 
$$A^n\lceil_{\S}=(\D^n+\beta\ov{\ihh})\lceil_{\S}\subset (\D^n+\beta\ov{\ihh})\lceil_{D(\D^n)},
$$
we obtain that 
$$A^n=(\D^n+\beta\ov{\ihh})\lceil_{D(\D^n)}.$$
Hence for $\beta$ with $|\beta|<1/C_{\rm I}$,  
$A^n$ is self-adjoint on $D\D^n$ and essentially self-adjoint on any core of $\D^n$.
Thus Proposition \ref{e2} (1) follows.
\qed

\bl{f7}
Let $g\in \S$ and $m\geq 1$. 
Then 
there exist 
$\g 1,...,\g {m-1}$, $\gk\in\S$, $j_1,...,j_\ell=1,...,\nu$, $\ell=1,...,m$, 
such that 
\eq{koo1}
\ada^m(g)=\sum_{\ell=0}^{m-1}\ada^\ell(\g \ell)+
\sum_{\ell=1}^m\J \gk \p 1\cdots \p \ell. 
\en 
\el
\proof 
We prove the lemma by  induction with respect to $m$. 
Let $m=1$. Then 
\eq{koo2}
\ada(g)=-g''+2 \sum_{j=1}^\nu g_j' p_j=
\ada^0(-g'')+2\sum_{j=1}^\nu g_j' p_j,
\en 
where 
$g''=\Delta g$ and $g_j'=-i\partial g/\partial x_j$. 
Thus \kak{koo1} follows for $m=1$. 
Suppose \kak{koo1}  holds for $m=0,1,...,k$. 
Then we have 
\eq{f8}
\ada^{k+1}(g)=\ada\ada^k(g)=
\sum_{\ell=0}^{k-1}\ada^{\ell+1}(\g \ell)+\sum_{\ell=1}^k \ada(\gk\p 1\cdots \p\ell).
\en 
Directly we can see by \kak{koo2} that 
\eqnn
&&
\ada(\gk \p 1\cdots \p \ell)=
\ada(\gk)\p 1\cdots \p \ell+\gk \ada(\p 1\cdots \p \ell)\nonumber \\
&&
=
(-(\gk)''+2\sum_{j=1}^\nu(\gk)'_j p_j)\p 1\cdots \p \ell+\gk[V,\p 1\cdots \p \ell]\nonumber \\
&&
=\gk\sum_{m=1}^{\ell-1}
\sum_{\{i_1,...,i_\ell\}\subset\{j_1,...,j_m\}}
V^{i_1,...,i_\ell}\underbrace{
\p1\cdots \sl{\pp 1}\cdots \sl{\pp \ell}\cdots \p m}_{\leq\ell-1}\nonumber \\
&&\label{f9} 
-(\gk)''\underbrace{\p 1\cdots \p \ell}_{\ell}
+ 2\sum_{j=1}^\nu (\gk)'_j \underbrace{p_j\p 1\cdots \p \ell}_{\ell+1}+
\underbrace{\gk V^{j_1,...,j_\ell}}_{=\ada^0(\gk V^{j_1,...,j_\ell})}.
\ennn 
Substituting \kak{f9} to \kak{f8} and rearranging, we can see that 
$$\ada^{k+1}(g)=\sum_{\ell=0}^{k}\ada^\ell(\tilde g^{(\ell)})+
\sum_{\ell=1}^{k+1}\J \tilde g_{j_1...j_\ell}^{(k)} \p 1\cdots \p \ell$$ 
with some $g^{(\ell)}$,  $\tilde g_{j_1...j_\ell}^{(k)}\in\S$, 
$j_1,...,j_\ell=1,..,\nu$, $\ell=1,...,k+1$. 
Thus the lemma follows.
\qed
\bl{f10}
Let $g\in \S$. Then there exists a constant $C_{g,m}$ such that 
\eq{koo3}
\|\ada^m(g)\Phi\|\leq C_{g,m}
(\|\D^{m/2}\Phi\|+\|\Phi\|),\ \ \ \Phi\in\S, \ \ \ m\geq 0.
\en 
\el
\proof 
We prove the lemma by  induction with respect to $m$.  
For $m=0$, \kak{koo3} follows. 
Assume that \kak{koo3} holds for $m=0,1,..,k$. Then 
by Lemma \ref{f7} we see that 
\eqn
 \|\ada^{k+1}(g)\Phi\|
&\leq &
\sum_{\ell=0}^k
\|\ada^\ell(\g \ell)\Phi\|+\sum_{\ell=1}^{k+1} \J\|\gk\p 1\cdots\p \ell\Phi\|\\
&\leq & 
\sum_{\ell=0}^{k}\|\ada^\ell(\g \ell)\Phi\|+
C \sum_{\ell=1}^{k+1}  \|\D^{\ell/2}\Phi\|\enn
with some constant $C$. 
By the assumption of the induction and \kak{f1} we have 
\eqn 
\|\ada^{k+1}(g)\Phi\|
&\leq& C'\sum_{\ell=0}^{k+1}(\|\D^{\ell/2}\Phi\|+\|\Phi\|)\\
&\leq & C''(\|\D^{(k+1)/2}\Phi\|+\|\Phi\|)
\enn 
with some constants $C'$ and $C''$. 
Then the lemma follows. 
\qed 
\bl{f11}
We have 
$\ada^k(p_j)=\beta \ada^{k-1}(i\partial V/\partial x_j)$ on $\S$.
\el
\proof 
We see that on $\S$, 
\eqn
\ada^k(p_j)
&=& 
\adxy^k(p_j)\\
&=& \adx^k(p_j)\\ 
&&+\beta \sum_j \underbrace{
\adx\cdots \stackrel{j}{\ady}\cdots \adx}_{k} (p_j)\\
&&+\beta^2 \sum_{j_1<j_2} \underbrace{
\adx\cdots \stackrel{j_1}{\ady}\cdots \stackrel{j_2}{\ady} 
\cdots \adx}_k  (p_j)\\
&&+\beta^3 \sum_{j_1<j_2<j_3}  
\underbrace{ 
\adx\cdots \stackrel{j_1}{\ady}\cdots \stackrel{j_2}{\ady} \cdots \stackrel{j_3}{\ady}
 \cdots \adx}_k (p_j)\\
&&\vdots \\
&&+\beta^k \ady^k(p_j).
\enn 
Since 
$\adx(p_j)=0$, 
we have 
\eqn
\ada^k(p_j)
&=&  \beta \underbrace{\adx\cdots \adx}_{k-1}  \ady (p_j)\\
&&+\beta^2 \sum_j \underbrace{
\adx\cdots \stackrel{j}{\ady}\cdots \adx \ady}_{k-1}(p_j)\\
&&+\beta^3 \sum_{j_1<j_2}  
\underbrace{
\adx\cdots \stackrel{j_1}{\ady}\cdots \stackrel{j_2}{\ady}
 \cdots \adx \ady}_{k-1}(p_j)\\
&&\vdots \\
&&+\beta^k \ady^{k-1}\ady (p_j)\\
&=& \beta \ada^{k-1}\ady(p_j)\\
&=& \beta \ada^{k-1}(i\partial V/\partial x_j).
\enn 
Thus the lemma follows. 
\qed
\bl{f12}
Let $k\geq 1$. 
Then there exists a constant $C_{V,k}$ such that 
$$\|\ada^k(p_j)\Phi\|\leq |\beta| C_{V,k}(\|\D^{(k-1)/2}\Phi\|+\|\Phi\|),\ \ \ \Phi\in\S.$$
\el
\proof 
By Lemmas \ref{f10} and \ref{f11} we have 
$$\|\ada^k(p_j)\Phi\|=|\beta| 
\|\ada^{k-1}(i\partial V/\partial x_j )\Phi\|\leq |\beta| C(\|\D^{(k-1)/2}\Phi\|+\|\Phi\|)$$
with some constant $C$. Then the lemma follows.
\qed
{\it Proof of Proposition \ref{e2} (2)}\\
Let $\Phi\in\S$. 
We have 
\eqn 
\|\D^n\Phi\|
&=& \|(\D^n+\beta\ihh-\beta\ihh)\Phi\|\leq \|A^n\Phi\|+|\beta|\|\ihh\Phi\| \\
&\leq& \|A^n\Phi\|+|\beta| C(\|\D^n\Phi\|+\|\Phi\|)
\enn
with some constant $C$. 
Hence it follows that 
$$\|\D^n\Phi\|\leq \frac{1}{1-|\beta| C}(\|A^n\Phi\|+\|\Phi\|)$$ 
for $\beta$ with $|\beta|<1/C$. 
Moreover
$$\|A^n\Phi\|\leq a\|\hat A^n\Phi\|+b\|\Phi\|,\ \ \ \Phi\in \S, $$
with some constants $a$ and $b$. 
Then 
\eq{e3}
\|\D^n\Phi\|\leq C'(\|\hat A^n\Phi\|+\|\Phi\|), \ \ \ \Phi\in \S, 
\en 
follows with some constant $C'$. Since $\S$ is  a core of  $\hat A ^n$, 
one can see that \kak{e3} can be extended to $\Phi\in D(\hat A^n)$.
Then 
\eq{e4}
\|\D^{n/2}\Phi\|\leq C''(\|\hat A^{n/2}\Phi\|+\|\Phi\|) ,\ \ \ \Phi\in D(\hat A^{n/2}), 
\en 
with some constant $C''$. 
By Lemma \ref{f12} and \kak{e4} it follows that 
$$\|\ada^k(p_j)\Phi\|\leq |\beta| C'''(\|\hat A^{(k-1)/2}\Phi\|+\|\Phi\|),\ \ \ \Phi\in\S,$$
with some constant $C'''$. 
In particular 
$$\|\ada^k(p_j)\Phi\|\leq |\beta| C(\|\hat A^{(k+1)/2}\Phi\|+\|\Phi\|)\ \ \ \Phi\in\S,$$
follows. 
Then the Proposition \ref{e2} (2) follows. 
\qed

{\it Acknowledgements.} 
{\footnotesize 
F.H. thanks  Grant-in-Aid  for Science Research (C) 15540191 from MEXT.
}

\end{document}